\documentclass[twocolumn]{revtex4-2}
\usepackage{graphicx}
\graphicspath{{figs/}}
\DeclareGraphicsExtensions{.pdf,.eps,.png,.jpg,.jpeg,.tif,.tiff}
\usepackage{amsmath, amssymb}
\usepackage{hyperref}
\usepackage{epstopdf}
\usepackage{float}
\usepackage{ragged2e}

\makeatletter
\long\def\@makecaption#1#2{%
  \vskip\abovecaptionskip
  \sbox\@tempboxa{#1: #2}%
  \ifdim \wd\@tempboxa >\hsize
    \hbox to \hsize{\parbox{\hsize}{\justifying\small#1: #2}}%
  \else
    \hbox to \hsize{\hfil\box\@tempboxa\hfil}%
  \fi
  \vskip\belowcaptionskip}
\makeatother

\begin{document}

\title{Machine Learning for Polymer Chemical Resistance to Organic Solvents}

\author{Shogo Kunieda$^{1}$, Mitsuru Yambe$^{1}$, Hiromori Murashima$^{1}$, Takeru Nakamura$^{1}$, 
Toshiaki Shintani$^{2}$, Hitoshi Kamijima$^{2}$, Yoshihiro Hayashi$^{3}$, 
Yosuke Hanawa$^{1}$ and Ryo Yoshida$^{3,4,5}$}

\affiliation{$^{1}$SCREEN Holdings Co., Ltd., Kyoto 602-8585, Japan}
\affiliation{$^{2}$Research Institute of Systems Planning, Inc., Tokyo 150-0031, Japan}
\affiliation{$^{3}$The Institute of Statistical Mathematics, Research Organization of Information and Systems, Tachikawa, Tokyo 190-8562, Japan}
\affiliation{$^{4}$Graduate Institute for Advanced Studies, SOKENDAI, Tachikawa, Tokyo 190-8562, Japan}
\affiliation{$^{5}$Advanced General Intelligence for Science Program (AGIS), RIKEN TRIP, Wako, Saitama 351-0198, Japan}

\maketitle

\onecolumngrid
\begin{quotation}
\noindent\textbf{Abstract.} Predicting the chemical resistance of polymers to organic solvents is a longstanding challenge in materials science, with significant implications for sustainable materials design and industrial applications.
Here, we address the need for interpretable and generalizable frameworks to understand and predict polymer chemical resistance beyond conventional solubility models.
We systematically analyze a large dataset of polymer--solvent combinations using a data-driven approach.
Our study reveals that polymer crystallinity and density, as well as solvent polarity, are key factors governing chemical resistance, and that these trends are consistent with established theoretical models.
These findings provide a foundation for rational screening and design of polymer materials with tailored chemical resistance, advancing both fundamental understanding and practical applications.
\end{quotation}

\vspace{1.5em}

\twocolumngrid

\section*{Introduction}

The compatibility between polymers and solvents represents a critical technological challenge across industry, academia, and government sectors,
including membrane separation, materials recycling, and drug delivery systems.
In membrane fabrication, the interaction between polymers and solvents has been widely recognized to strongly influence
phase separation behavior and membrane morphology, and practical solvent selection is often guided by parameters
such as Hansen Solubility Parameters (HSP).\cite{Hansen1967,Hansen2007}
In solvent-based recycling, computational chemistry is increasingly employed to predict polymer--solvent dissolution behavior,
aiming to establish highly efficient separation and recovery processes based on selective dissolution.\cite{Zhou2023}
In the field of drug delivery, the solubility and swelling properties of polymer carriers are key factors in controlling
drug release profiles and targeting capability.
Advanced drug delivery systems based on stimuli-responsive polymers or self-assembled micelles have been actively explored
for these purposes.\cite{Peppas2006,Zhou2018}
A common feature across these applications is that polymer--solvent compatibility is governed by a complex interplay
of chemical factors such as molecular structure, polarity, molecular weight, and branching.

In this study, we focus on the chemical resistance of polymers.
Chemical resistance is a critical property for ensuring safety and long-term reliability in solvent-exposed environments
such as pipes, tanks, and gaskets in chemical plants, as well as in analytical instruments and microfluidic devices.
Fluoropolymers, including Per- and polyfluoroalkyl substances (PFAS)-based materials such as polytetrafluoroethylene (PTFE) and fluorinated ethylene propylene (FEP), have traditionally been employed in such applications
due to their excellent resistance properties.
However, concerns over their environmental persistence and toxicity have led to increasing international regulatory pressure
on the use of PFAS compounds.\cite{Blum2015,Brown2020}
As a result, efforts to phase out non-essential uses of PFAS are accelerating under the "essential use" concept.\cite{Cousins2019}
creating a strong demand for the development of polymeric materials that offer both high chemical resistance
and improved environmental degradability.
Against this backdrop, the ability to quantitatively predict and understand chemical resistance is of paramount importance
for the design of sustainable materials.

To evaluate the compatibility between polymers and solvents, several theoretical approaches have been developed,
including the Hildebrand solubility parameter ($\delta$), HSP,
and the Flory--Huggins interaction parameter ($\chi$).
The Hildebrand parameter, one of the earliest theories, defines solubility as a one-dimensional quantity
based on the cohesive energy density, which quantifies intermolecular attractions.
This parameter serves as a numerical representation of the empirical rule "like dissolves like".\cite{Hildebrand1950}
Building upon this, Hansen introduced the HSP theory, which decomposes the total solubility parameter into three components---
dispersion, polarity, and hydrogen bonding---and evaluates compatibility based on the distance between polymer and solvent
in the three-dimensional solubility space.\cite{Hansen1967,Hansen2007}
Meanwhile, the Flory--Huggins theory has been widely employed as a thermodynamic framework grounded in statistical mechanics,
in which the mixing free energy of a polymer--solvent system is expressed using the $\chi$ parameter.\cite{Flory1942,Huggins1941}
These theories have served as fundamental tools in understanding the physicochemical behavior and solubility of polymers,
contributing to progress in diverse applications including materials design, membrane formation, drug delivery,
and solvent-based recycling.

In recent years, data-driven approaches known as materials informatics have emerged as a powerful extension
to these traditional physicochemical frameworks.\cite{Kim2018,Xu2020,Aoki2023,Yu2023,Kuenneth2023,Agarwal2025}

Kim et al. developed Polymer Genome, a machine learning platform capable of accurately predicting seven polymer properties---
including glass transition temperature, solubility parameters, and density---
using 229-dimensional structural descriptors as input features.\cite{Kim2018}
Xu et al. constructed a kernel regression model using solvent properties and Hansen solubility parameter-based descriptors,
and identified key governing factors such as polarity and molecular size for specific polymers (e.g., polydimethylsiloxane (PDMS)).\cite{Xu2020}
Aoki et al. proposed a multitask learning framework that integrates experimental data and quantum chemical calculations
to simultaneously predict Flory--Huggins $\chi$ parameters and solubility labels.\cite{Aoki2023}
Their model demonstrated high generalization performance and extracted latent features not captured by traditional theories.
Yu et al. applied a natural language processing model, SolvBERT, which takes the chemical structures of solutes and solvents
as input to predict solvation free energies and solubility, achieving performance comparable to conventional descriptor-based approaches.\cite{Yu2023}
Kuenneth et al. developed an end-to-end machine learning pipeline by training a transformer-based language model, polyBERT,
on over one million hypothetical polymer structures, enabling rapid and accurate prediction of 29 different polymer properties.\cite{Kuenneth2023}
Agarwal et al. fine-tuned GPT-3.5 to classify the solubility of polymer--solvent pairs in a binary Yes/No format,
achieving intuitive and accurate predictions (Yes accuracy: 0.90; No accuracy: 0.83) without the need for descriptor engineering
or complex preprocessing.\cite{Agarwal2025}
Recent advances in materials informatics have enabled high-accuracy predictions of polymer properties and solubility
using machine learning techniques.
However, most previous studies have focused on predicting solubility and compatibility between polymers and solvents,
whereas the prediction and analysis of chemical resistance of polymers against organic solvents has been scarcely explored.
This study aims to address this gap by developing a machine learning framework specifically for chemical resistance prediction.

In this work, we construct a machine learning framework to predict chemical resistance using a curated dataset
of over 2,200 polymer--solvent combinations.
The model incorporates molecular descriptors derived from MD simulations
and quantum chemical calculations, allowing for generalizable and interpretable classification.
Moreover, by analyzing the contribution of each descriptor to the prediction outcomes with established chemical insights,
this study seeks to deepen the understanding of polymer--solvent compatibility and establish a data-driven yet interpretable framework
that supports the rational design of polymer materials.

\begin{figure*}[t]
    \centering
    \includegraphics[width=\linewidth]{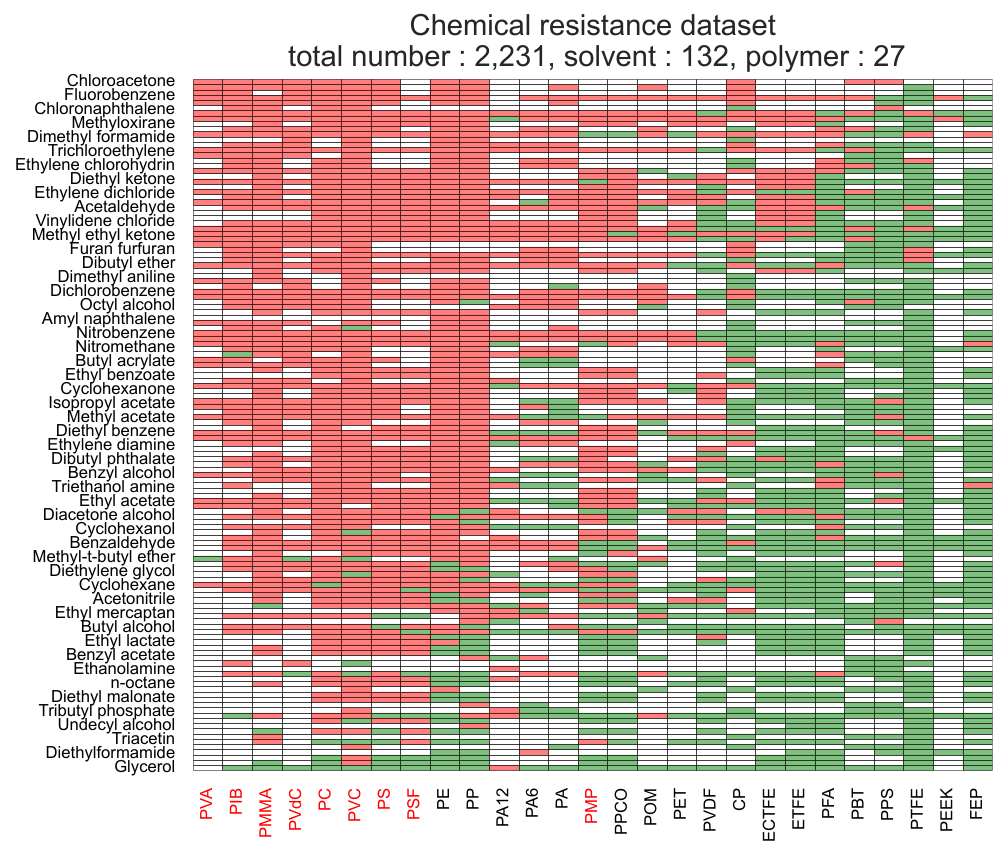}
    \caption{Overview of the chemical resistance dataset. The horizontal axis represents 27 polymers, and the vertical axis represents 132 organic solvents. Polymers are categorized as either crystalline (black) or amorphous (red). Each cell indicates the presence or absence of chemical resistance for a given polymer–solvent pair (green: resistant, red: non-resistant, white: no data). The dataset contains chemical resistance information for 2,231 polymer–solvent combinations.}
    \label{fig:overview}
\end{figure*}

\begin{figure*}[t]
    \centering
    \includegraphics[width=\linewidth]{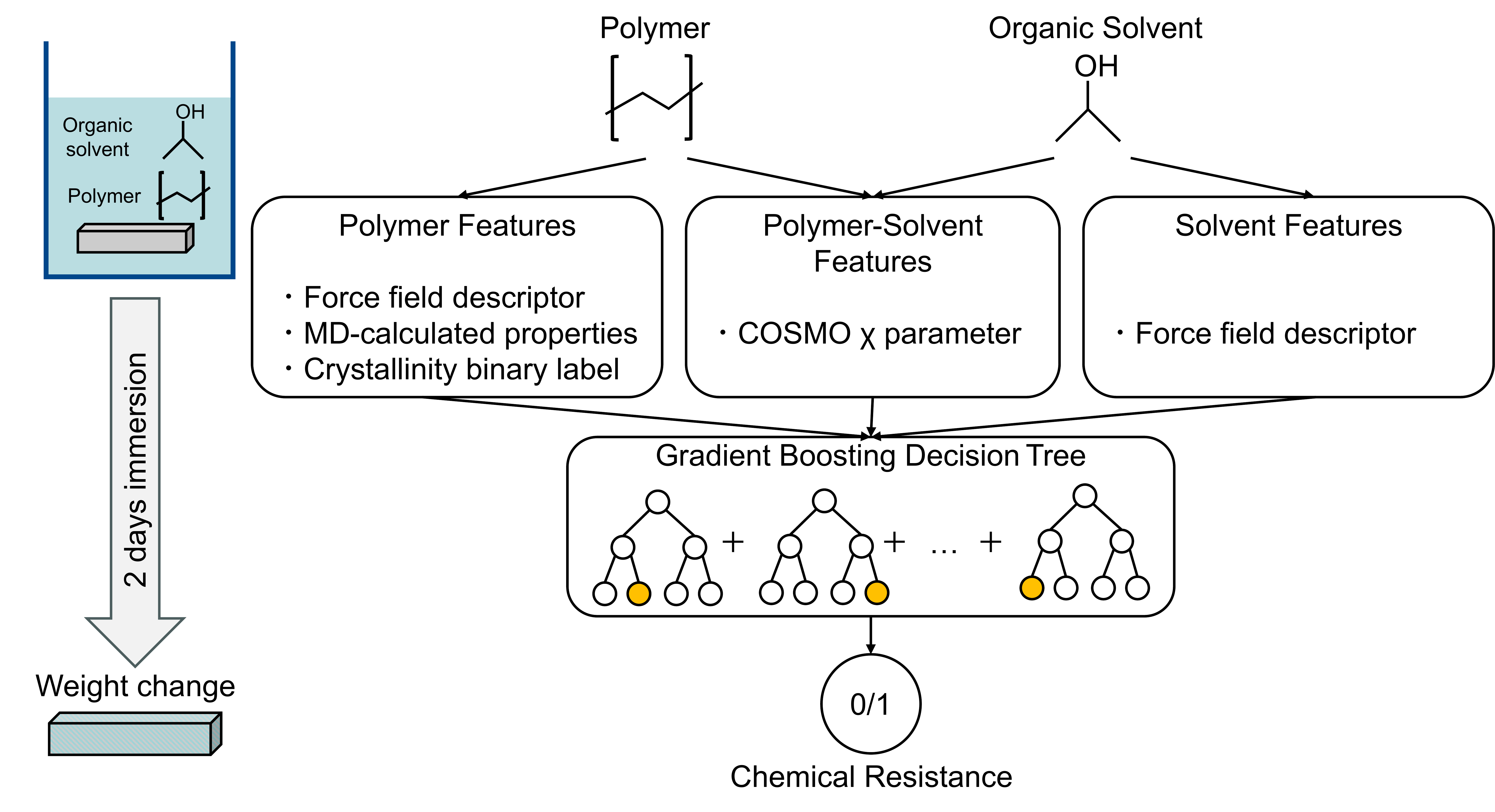}
    \caption{Overview of features, model, and target used for predicting polymer chemical resistance. Features include the GAFF2-based force field descriptor, MD properties from RadonPy, crystallinity label, and COSMO-RS $\chi$ parameter. A gradient boosting decision tree classifier was used to predict binary resistance labels.}
    \label{fig:dataset}
\end{figure*}

\section*{Methods}

\subsection*{Datasets}
To construct a chemical resistance dataset for machine learning-based prediction, data were collected and merged
from multiple manufacturers' reports on chemical resistance.\cite{Kayo,ASOH,PISCOCorporation,VICIAGInternational,ProfessionalPlastics,ThermoFisherScientific}
Although each source provides information on the chemical resistance of polymer--solvent combinations,
the evaluation criteria vary across datasets.
In this study, we merged multiple datasets with the aim of systematically analyzing the chemical resistance
of various resins against organic solvents.
We define the binary label as y=1 for non-resistant and y=0 for resistant.
The model outputs \( p := P(y=1 \mid x) \), i.e., the predicted probability of non-resistance.
Unless otherwise noted, 'probability' refers to \( P(y=1 \mid x) \).
A binary chemical resistance label was defined as follows: combinations meeting the most stringent resistance criterion
within a given dataset were labeled as 0 (resistant), and all others were labeled as 1 (non-resistant).
When multiple resistance ratings existed for a given polymer--solvent pair across different sources,
a binary value of 0 was assigned to the pair only if all sources labeled it as resistant;
otherwise, a value of 1 was assigned.

As a result of this merging process, the final dataset consisted of the 2,231 combinations
27 polymers and 132 organic solvents (Figure 1).
The 132 solvents were sorted based on the proportion of non-resistant entries and clustered into ten groups,
which were then used for model validation via leave-one-group-out cross-validation (LOGOCV) as detailed later.
The distribution of polymer data is shown in Figure S1, and the clustering of solvents is illustrated in Figure S2.
For the polymer property analysis in the Polymer crystallinity and Polymer features subsections, a total of 40,971 polymers with Simplified Molecular Input Line Entry System (SMILES) representations were used,
derived from the fully automated molecular dynamics library RadonPy.\cite{Hayashi2022}
For the solvent feature analysis in the Solvent features subsection and the solubility parameter analysis in the Solubility Parameter subsection,
we used 9,828 solvents included in the Hansen Solubility Parameters in Practice (HSPiP) dataset, along with their corresponding SMILES representations.\cite{HSP}

\subsection*{Descriptors}

In this study, four different methods were used to generate feature representations for the model.

\textbf{Force field descriptor:}
The force field descriptor vectorizes the compositional and structural features of a given polymer or organic solvent 
based on an empirical potential used for MD simulations.\cite{Kusaba2023}
The AMBER (GAFF2) potential includes ten distinct types of force field parameters, e.g., atomic mass, 
depth of the potential well of van der Waals interactions, force constant of bond stretching, 
that are defined for their respective molecular components, such as atoms, bonds, bond angles, and dihedral angles.\cite{Wang2004}
The force field descriptor encodes the distribution (i.e., a histogram or occurrence frequency) of parameter values 
for their molecular component into a 190-dimensional fixed-length vector, using a machine learning technique called the kernel mean embedding.\cite{Muandet2017}
Descriptor calculation was performed using the Python library RadonPy.\cite{RadonPyGithub}
For solvents, descriptors were calculated directly from single SMILES strings, while for polymers, 
a cyclic SMILES representation was constructed by connecting 10 repeating units end-to-end.
Further computational details are provided in the Supporting Information.

\textbf{$\chi$ parameter:}
To characterize the solubility between polymers and solvents, the Conductor-like Screening Model for Real Solvents (COSMO-RS) $\chi$ parameter was calculated.
COSMO files were generated using TURBOMOLE,\cite{TURBOMOLE} and $\chi$ parameter calculations were performed using COSMOtherm.\cite{COSMOthermo}
Details of the computational procedure are available in the Supporting Information.

\textbf{MD parameters:}
Equilibrium MD simulation results were used as explanatory variables.
Sixteen polymer properties---namely density, radius of gyration ($R_g$), self-diffusion coefficient, heat capacities ($C_p$ and $C_v$), 
compressibility, isentropic compressibility, bulk modulus, isentropic bulk modulus, 
volume expansion coefficient, linear expansion coefficient, mean square displacement ($r^2$), 
static dielectric constant, DC dielectric constant, nematic order parameter, 
and refractive index---were obtained from the RadonPy dataset and used as features.

\textbf{Crystallinity label:}
The 27 polymers in the chemical resistance dataset were labeled as either crystalline or amorphous.
A classification model trained using the force field descriptors and MD parameters of these polymers was used 
to predict crystallinity for virtual polymers in the extended dataset.

\begin{figure*}[t]
  \centering
  \includegraphics[width=\linewidth]{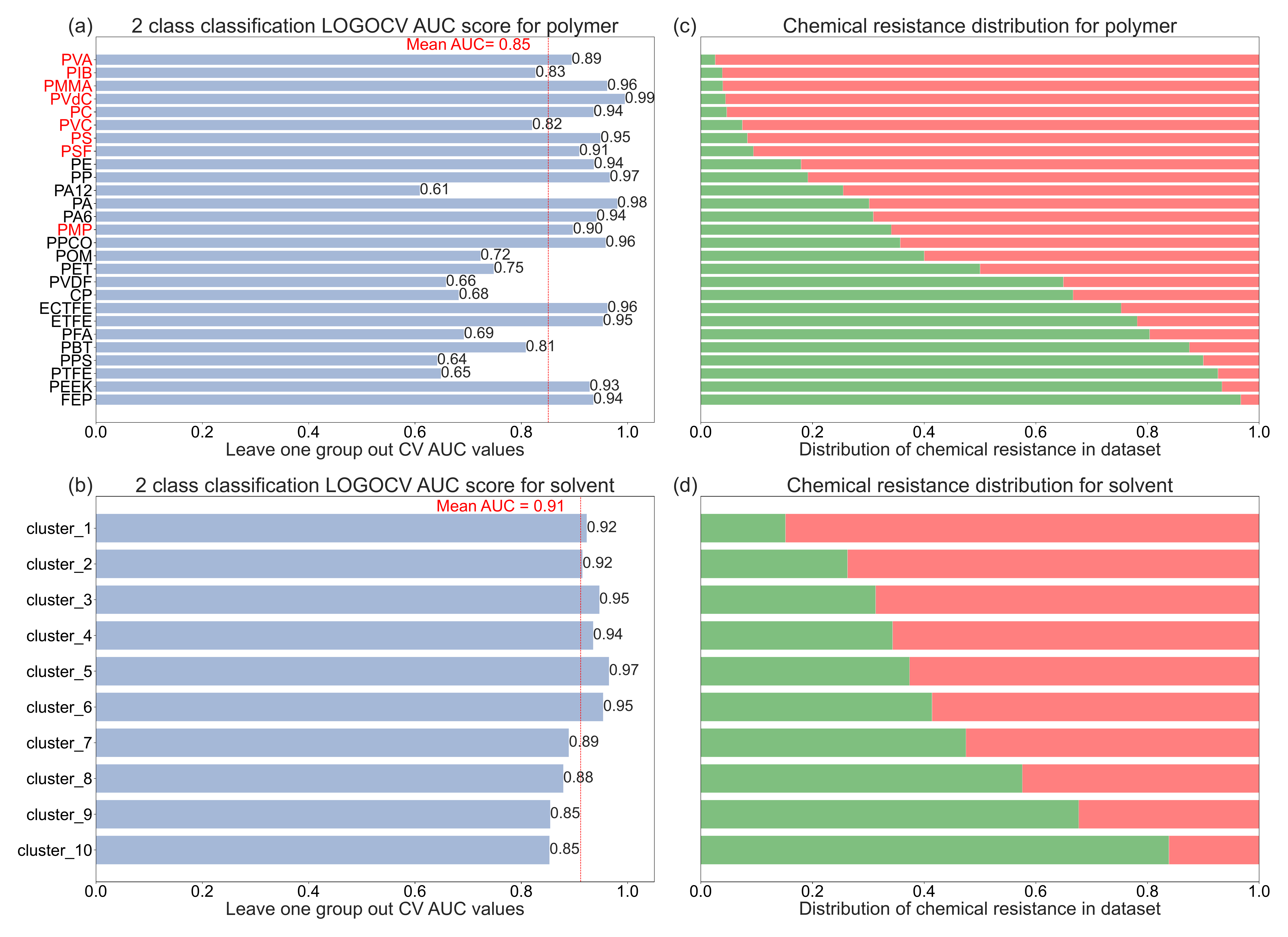}
  \caption{Model validation performance and data distribution visualization for the chemical resistance dataset.
  (a) LOGOCV ROC-AUC evaluation, with each polymer treated as a separate group.
  (b) Proportion of chemical resistance for each polymer (x-axis: proportion of resistant cases; green: resistant, red: non-resistant). 
  Polymers are sorted in ascending order of resistance.(c) LOGOCV ROC-AUC evaluation, with each solvent cluster treated as a group.(d) Proportion of chemical resistance for each solvent (x-axis: proportion of resistant cases; green: resistant, red: non-resistant). Solvents are sorted in ascending order of resistance.}
  \label{fig:features}
\end{figure*}

\subsection*{Model}

Figure 2 illustrates the data flow into the chemical resistance prediction model.
The features used include the previously described force field descriptor, COSMO-RS $\chi$ parameter, MD-calculated properties, and crystallinity binary label.
For the machine learning algorithm, a gradient boosting decision tree implemented in the scikit-learn library was employed.\cite{Pedregosa}
The model outputs a probability $p$ representing $P(y=1|x)$, the probability of non-resistance given the input features $x$.
Model performance was evaluated using the area under the receiver operating characteristic curve (ROC-AUC) based on LOGOCV across 27 polymers and 10 solvent clusters.

\section*{Results and Discussion}

\subsection*{Model accuracy}

Figure 3 presents the predictive performance of the machine learning model constructed using the chemical resistance dataset comprising the 2,231 entries. 
Figure 3a shows the LOGOCV ROC-AUC results, 
where cross-validation was performed for each polymer cluster treated as a separate group. 
Figure 3b displays the proportion of resistant cases for each polymer, 
calculated directly from the ground truth labels in the 2,231 experimental data points, 
arranged in descending order of resistance---i.e., polymers with higher resistance 
(higher proportion of resistant cases) appear at the top.

\begin{figure*}[t]
    \centering
    \includegraphics[width=\linewidth]{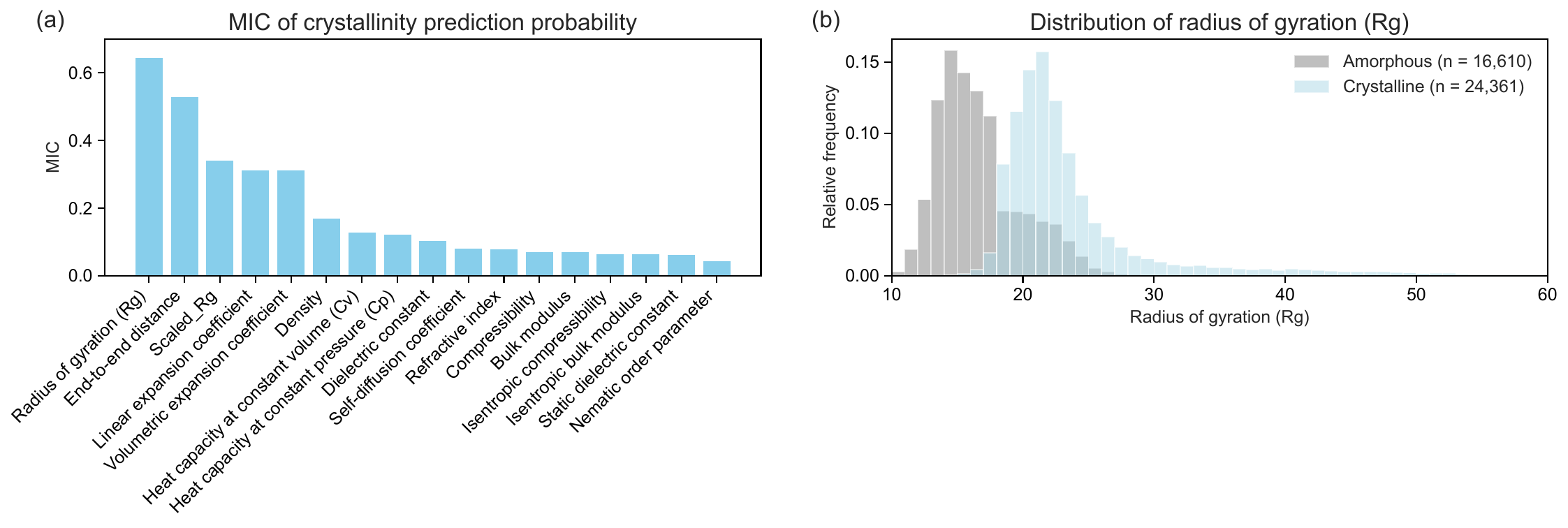}
    \caption{Feature importance and property distributions related to polymer crystallinity.(a) Bar chart of the maximal information coefficient (MIC) between predicted crystallinity probabilities and polymer properties. The model was trained using crystallinity labels for 27 real polymers and applied to 40,971 virtual polymers. Bars are ordered from left to right by decreasing MIC.(b) Histogram of the radius of gyration ($R_g$) for the virtual polymers. Polymers predicted as crystalline are shown in light blue, and amorphous ones in gray.}
    \label{fig:validation}
\end{figure*}

Similarly, Figure 3c shows the LOGOCV ROC-AUC results for solvent clusters, while Figure 3d illustrates the proportion of resistance labels within each cluster. 
Solvent clusters are sorted from top to bottom in increasing order of resistance, meaning clusters with higher solubility (lower resistance) are shown first. 
The definition of solvent clusters is provided in Figure S2.
All model evaluations are based on LOGOCV. On average, the model achieved a ROC-AUC of 0.85 at the polymer level and 0.91 at the solvent cluster level. 
An ROC-AUC of 0.85 indicates strong classification performance, while a value of 0.91 represents excellent discriminative ability, 
indicating strong classification performance overall.\cite{Liu2020} Detailed results, including confusion matrices and ROC curves for each cross-validation, 
are presented in Figure S3. It is noteworthy that in Figure 3a, a few polymers---particularly polyphenylene sulfide (PPS)---exhibited relatively low ROC-AUC values. 
This may be attributed to the limited presence of structurally similar polymers in the dataset, 
which likely reduced the model’s ability to generalize predictions for PPS  
Despite some limitations in prediction performance for specific polymers, the results demonstrate that 
a machine learning model incorporating relatively simple descriptors can achieve reasonable classification accuracy, 
even with a dataset of modest size. 
This finding underscores the potential of data-driven approaches for chemical resistance prediction.

\subsection*{Polymer crystallinity}

Figure 4 presents the analysis of feature importance and property distributions related to polymer crystallinity. 
In Figure 4a, the maximal information coefficient (MIC) was calculated between the predicted crystallinity probabilities from the classification model 
and various polymer properties.\cite{Reshef2011} MIC is a measure of the strength of association 
between two variables that can capture both linear and non-linear relationships, 
with values ranging from 0 (no association) to 1 (perfect association). 
This crystallinity model was trained using crystallinity labels for 27 polymers, 
and the model’s performance is shown in Figure S4. Due to the limited dataset size, 
the results should be interpreted as indicative trends rather than definitive conclusions.

Applying this model to 40,971 virtual polymers from the RadonPy property database resulted in approximately 24,361 polymers classified as crystalline 
and 16,610 as amorphous. Based on these predictions, MIC values were computed between crystallinity and MD-derived polymer properties. 
The radius of gyration ($R_g$) and end-to-end distance ($\langle R^2 \rangle$) showed the strongest correlations.

\begin{figure*}[t]
    \centering
    \includegraphics[width=\linewidth]{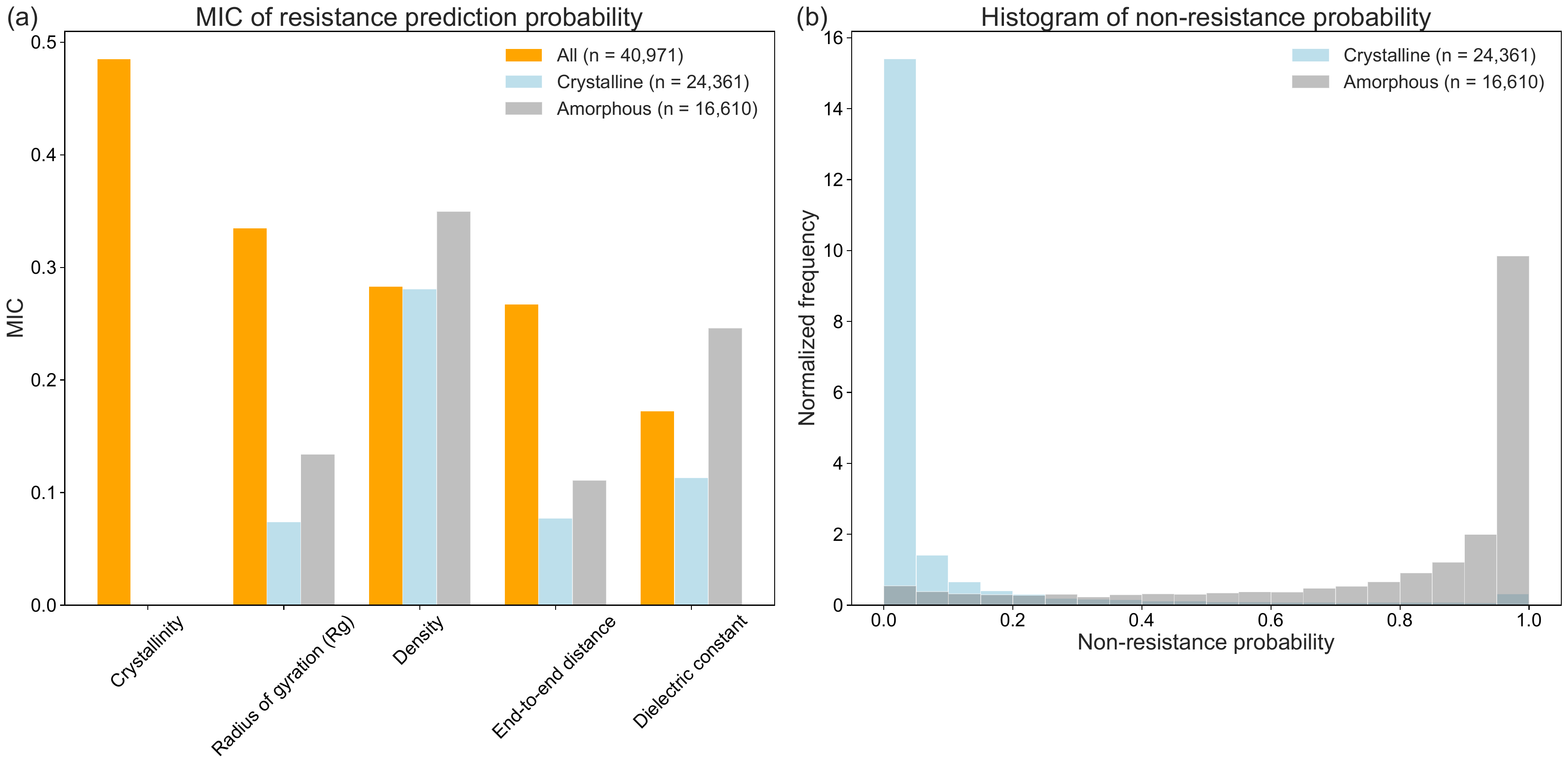}
    \caption{Prediction of chemical resistance and feature importance analysis based on polymer descriptors.(a) Correlation between predicted non-resistance probabilities $P(y=1|x)$ and polymer features, evaluated using the MIC. Orange, light blue, and gray bars represent MIC values for all polymers, crystalline polymers, and amorphous polymers, respectively. Features are ordered on the x-axis in descending order of MIC.(b) Histogram of $P(y=1|x)$ for 40,971 virtual polymers against MPK, as predicted by the model. The x-axis represents the predicted probability of non-resistance, and the y-axis shows the number of corresponding data points. Light blue and gray indicate polymers classified as crystalline and amorphous, respectively. Crystallinity was assigned using the crystallinity classification model trained on 27 known polymers.}
    \label{fig:crystallinity}
\end{figure*}

Figure 4b displays a histogram of $R_g$, with crystalline polymers shown in light blue and amorphous polymers in gray. 
The data reveal that crystalline polymers tend to exhibit larger $R_g$ values compared to amorphous ones. A similar trend was also observed for $\langle R^2 \rangle$.

A larger $R_g$ indicates that a polymer chain tends to adopt an extended conformation, whereas a smaller $R_g$ implies limited chain extension. 
Therefore, crystalline polymers are more likely to assume elongated conformations that facilitate alignment with neighboring chains, promoting crystallization. 
In contrast, amorphous polymers tend to adopt compact structures, hindering the formation of ordered crystalline regions.

This interpretation is broadly supported by previous MD simulation studies. 
For instance, it has been reported that reduced chain entanglement---characterized by increased chain mobility 
and a larger $R_g$---facilitates crystallization.\cite{Luo2013,Luo2016,Luo2016a} Additionally, 
polymer chains with linear structures (i.e., minimal branching or reduced influence from chain ends) have been shown to more readily form crystalline nuclei.\cite{Xiao2017,Yamamoto2019}
Drawing-induced increases in $R_g$ have also been associated with the formation of aligned, fibrous crystalline structures.\cite{Yamamoto2019}

Collectively, these studies demonstrate that crystallization behavior is highly dependent on topological factors such as the initial conformation, orientation, mobility, 
and entanglement state of polymer chains.

More recently, data-driven approaches using machine learning have enabled the classification of crystalline versus amorphous phases, 
and the identification of early signs of structural ordering, based on local structural fingerprints derived from MD simulations.\cite{Zhang2018} In such analyses, 
structural features reflecting chain orientation and linearity tend to emerge as key discriminative indicators.

Therefore, the strong correlation observed in this study between MD-derived physical descriptors 
and predicted crystallinity is in good agreement with the findings of previous studies.

It is also worth noting that the MD library used in this work, RadonPy, performs all-atom simulations based on initially amorphous structures.\cite{Hayashi2022} 
Despite this, the resulting physical descriptors still showed strong alignment with experimentally determined crystallinity classifications. 
This intriguing outcome reveals that inherent crystallization tendencies may be embedded within the computed physical properties themselves.

\begin{figure*}[t]
    \centering
    \includegraphics[width=\linewidth]{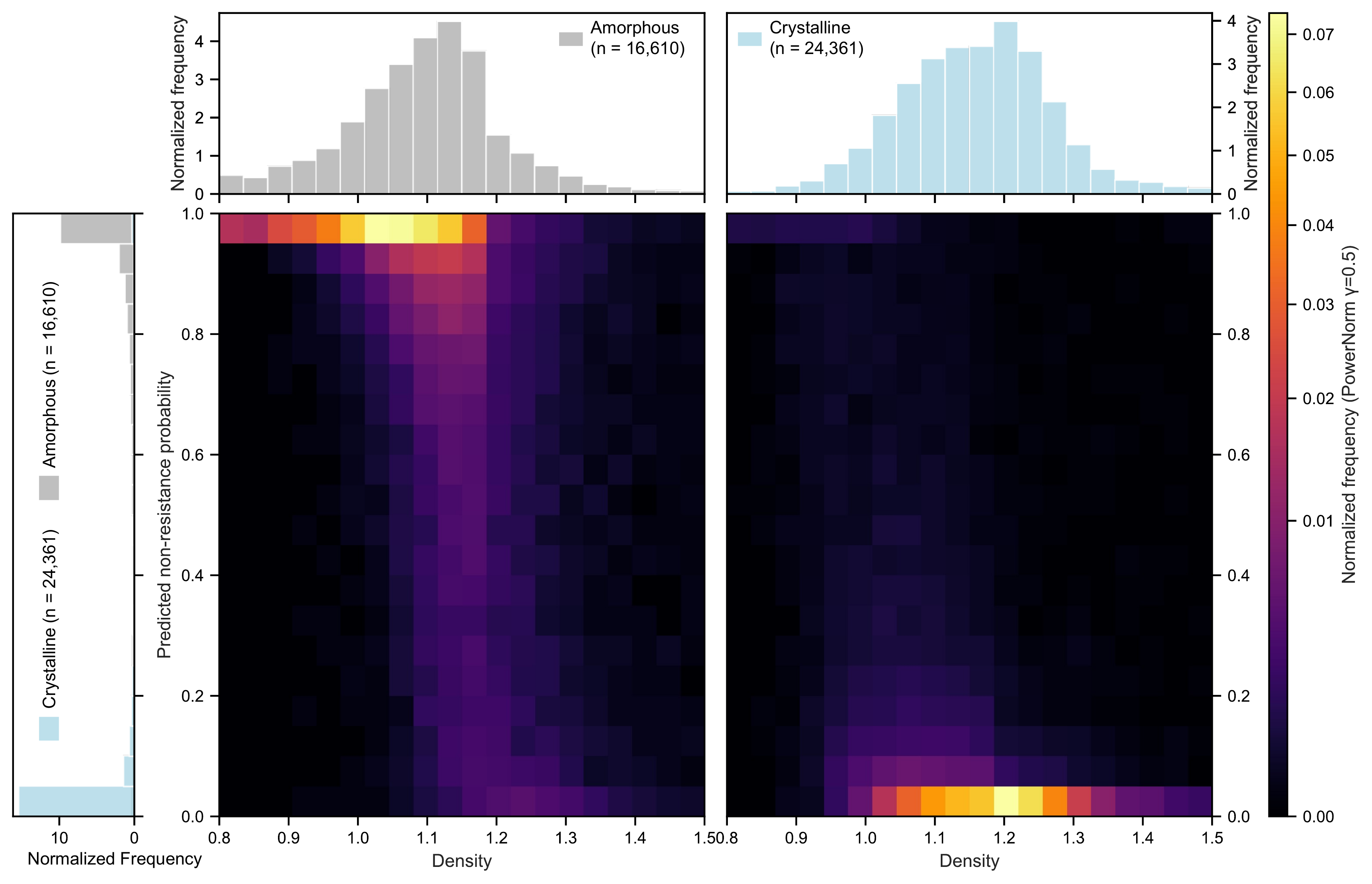}
    \caption{Two-dimensional heatmap showing the relationship between polymer density and predicted non-resistance probability $P(y=1|x)$. The x-axis represents polymer density, and the y-axis indicates the predicted probability of non-resistance against methyl propyl ketone (MPK). Predictions were made for 40,971 virtual polymers using the model trained on 2,231 experimental data points. The heatmap reflects the density of polymers in each region, with marginal histograms for non-resistance probability (top) and density (left). In the histograms, crystalline and amorphous polymers are shown in light blue and gray, respectively.}
    \label{fig:polymer_features}
\end{figure*}

\subsection*{Polymer features}

\begin{figure*}[t]
    \centering
    \includegraphics[width=\linewidth]{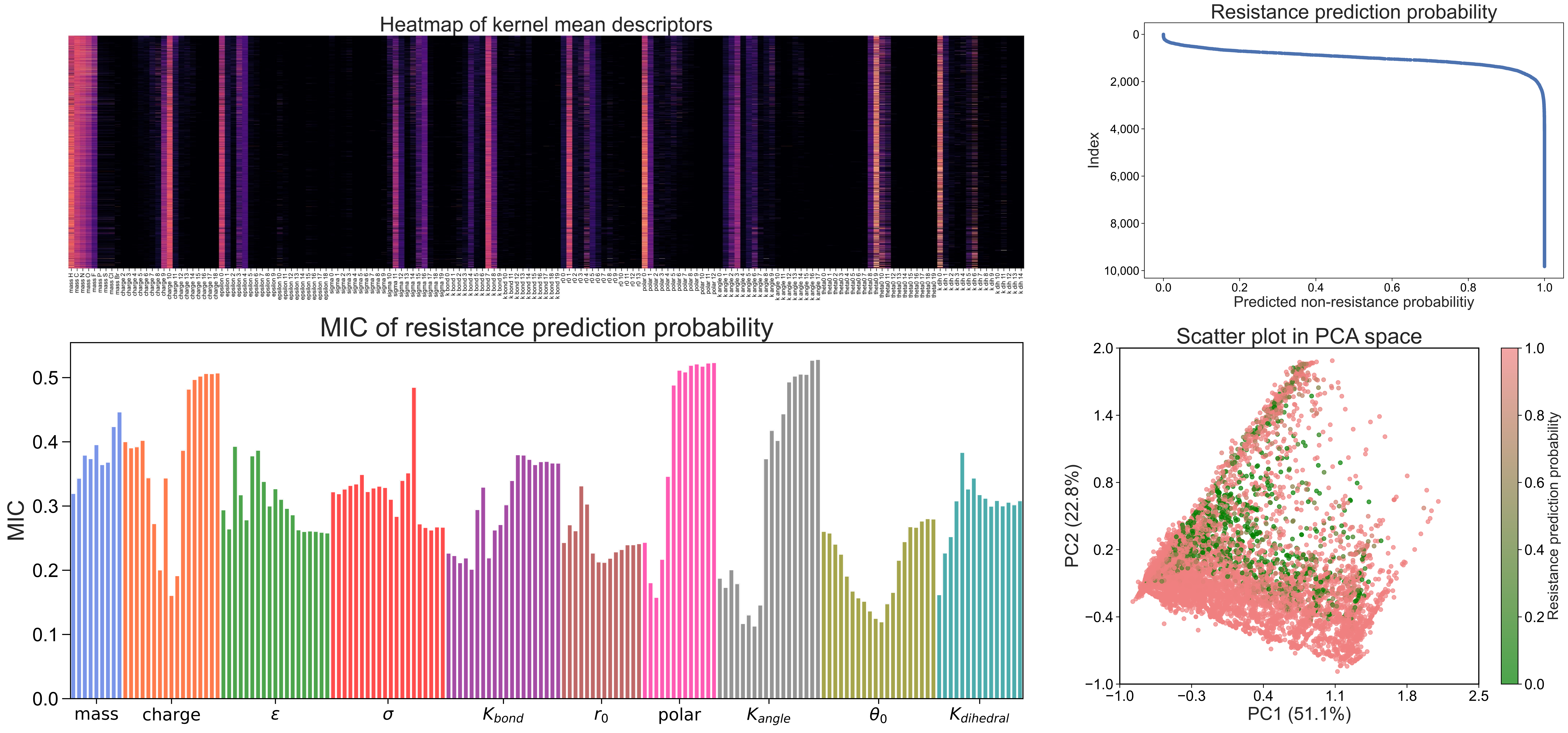}
    \caption{Prediction and visualization of chemical resistance descriptors for 9,828 organic solvents against PE.
(a) Heatmap of kernel mean descriptors. Solvents are sorted from top to bottom in ascending order of predicted non-resistance probability $P(y=1|x)$ against PE. The descriptors include atomic mass, atomic charge, van der Waals potential depth, equilibrium interaction distance, bond force constant, bond equilibrium length, bond polarity, angle force constant, equilibrium bond angle, and dihedral rotational barrier (10 parameters total). (b) Predicted probability of non-resistance $P(y=1|x)$ on the x-axis, with solvents indexed on the y-axis in ascending order of non-resistance probability. (c) MIC between non-resistance probability and each descriptor. Features correspond to those shown in the heatmap in (a). (d) Two-dimensional visualization of solvent descriptors by principal component analysis (PCA). The x- and y-axes represent the first and second principal components, respectively, and the color of each point indicates the predicted non-resistance probability $P(y=1|x)$.}
    \label{fig:density_resistance}
\end{figure*}

\begin{figure*}[t]
    \centering
    \includegraphics[width=\linewidth]{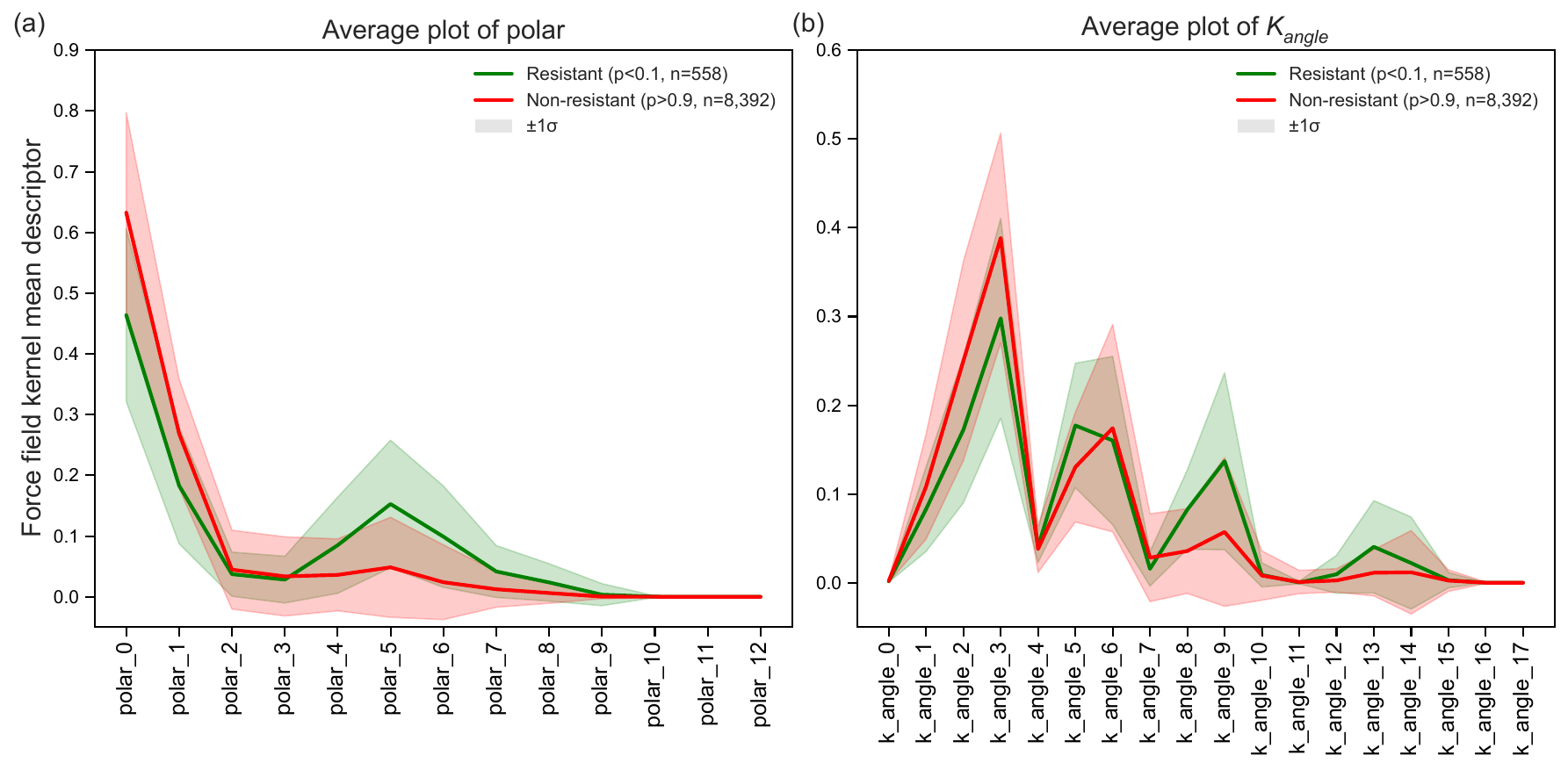}
    \caption{Comparison of data distributions for force field parameters based on predicted non-resistance probabilities $P(y=1|x)$.
Here, $p = P(y=1 \mid x)$ denotes the non-resistance probability.
(a) polarity, (b) bond angle force constant ($K_{\text{angle}}$). In both plots, data points with predicted non-resistance probabilities $> 0.9$ are shown in red, and those $< 0.1$ are shown in green. The x-axis represents the value of the corresponding force field parameter, and the y-axis indicates its magnitude. Solid lines show the mean values for each group, and the shaded areas represent standard deviations. Distributions for the remaining eight parameters are provided in Figure S5.}
    \label{fig:solvent_features}
\end{figure*}

Figure 5 presents the analysis of polymer properties in relation to chemical resistance prediction based on the trained model. 
In panel (a), the MIC was calculated between each polymer feature and the predicted probability of non-resistance $P(y=1|x)$. 
Among all features, the crystallinity classification label (0/1) exhibited the highest MIC value of 0.485, 
indicating a strong correlation with the predicted non-resistance probability $P(y=1|x)$.

Panel (b) shows histograms of the predicted non-resistance probabilities $P(y=1|x)$ stratified by crystallinity. 
For crystalline polymers, the predicted non-resistance probabilities $P(y=1|x)$ are concentrated near 0 (indicating high resistance),
while for amorphous polymers, they are skewed toward 1 (indicating low resistance). 
These distributions suggest that the model relies heavily on the crystallinity label to predict chemical resistance.

Since the predicted crystallinity probability was strongly correlated with $R_g$,
$R_g$ also exhibited a high MIC value in Figure 5a. 
Furthermore, Figure 5a presents stratified MIC evaluations between predicted non-resistance probabilities $P(y=1|x)$ and each feature, 
separately for the 24,361 crystalline and 16,610 amorphous polymers. In both groups, density showed the highest correlation with resistance prediction.

These results suggest that, in addition to crystallinity, polymer density may significantly influence chemical resistance prediction. 
To quantitatively examine this relationship, a two-dimensional heatmap of polymer density and predicted non-resistance probability $P(y=1|x)$ is shown in Figure 6.

Both categories exhibit a clear trend: as polymer density increases, the predicted probability of non-resistance $P(y=1|x)$ decreases, 
indicating that higher-density polymers are generally predicted to be more resistant.

This observation agrees with prior studies. For example, in semicrystalline PE, 
an increase in crystallinity---reflected by higher density---has been shown to reduce solvent diffusion coefficients, 
as the crystalline domains act as physical barriers to solvent penetration.\cite{Lutzow1999} Similarly, 
in thermally treated or crosslinked polyimides and block copolymer membranes, 
the formation of dense network structures leads to a reduction in free volume between polymer chains, 
thereby suppressing solvent uptake and swelling.\cite{Xu2021,Rangou2022}

Even in the absence of chemical crosslinking, polyimides incorporating rigid isohexide backbones have been reported to exhibit reduced segmental mobility 
and lower free volume, contributing to improved thermal and structural stability.\cite{Ji2015}

These findings suggest that factors such as crosslinking density and molecular packing can play a central role in governing solvent resistance---beyond 
the simple presence or absence of crystalline domains.\cite{Du2013} 
For instance, in phase-separated amorphous copolymer membranes, differences in crosslinking density have been shown to significantly influence morphological integrity 
and transport properties in solvent environments.\cite{Dugas2023}

However, it should be noted that the stratified MIC values from Figure 5a were 0.281 for crystalline polymers and 0.350 for amorphous ones, 
suggesting only moderate or lower correlations with density. This indicates that chemical resistance cannot be fully explained by polymer properties alone. 
Therefore, in the next section, we explore the contribution of solvent-side descriptors to resistance prediction.

\subsection*{Solvent features}

Figure 7 illustrates the relationship between predicted non-resistance probabilities $P(y=1|x)$ and solvent descriptors for 9,828 organic solvents against PE. 
The solvent descriptors are based on the FFKM descriptor. 
FFKM uses ten parameters derived from the GAFF2 force field---atomic mass, atomic charge, van der Waals potential depth, 
equilibrium interaction distance, bond stretching force constant, bond equilibrium length, bond polarity, angle bending force constant, equilibrium bond angle, 
and dihedral rotational barrier height. For each parameter, FFKM constructs atom- and bond-level probability density functions 
and averages them to obtain a feature vector representing the entire molecular structure. Further details are provided in Supporting Information S2.

In Figure 7a, the heatmap shows the relationship between FFKM descriptor values (x-axis) 
and solvents sorted by ascending predicted non-resistance probability $P(y=1|x)$ (y-axis). 
This visualization reveals patterns in the descriptors that correspond to chemical resistance levels, 
suggesting the presence of solvent groups that are distinguishable based on physical property trends. 
Figure 7b plots the predicted non-resistance probabilities $P(y=1|x)$, with solvents sorted accordingly. 
While nearly all of the 9,828 solvents are predicted to be non-resistant, 
the high validation performance observed for solvent clusters (ROC-AUC = 0.91, as shown in Figure.3(c)) confirms that 
the FFKM descriptors possess strong discriminative power.

In Figure 7c, the MIC was computed between each descriptor and the predicted non-resistance probability $P(y=1|x)$, 
corresponding to the descriptors shown in the heatmap in Figure 7a. Among the ten descriptors, charge, polar, 
and $K_{\text{angle}}$ (angle bending force constant) exhibited the highest correlations with predicted resistance. 
Figure 7d projects the prediction results for all 9,828 solvents onto a two-dimensional space defined by the first and second principal components of the solvent descriptors.
The triangular distribution pattern observed in the principal component analysis (PCA) plot is consistent with prior studies on force field-based solvent descriptor spaces.\cite{Kusaba2023}

Figure 8a presents an analysis of the polar (polarity) parameter, 
showing the mean and standard deviation of polarity values for two groups of solvents stratified by predicted non-resistance probability $P(y=1|x)$ 
($\geq$0.9: non-resistant (low resistance), $\leq$0.1: resistant (high resistance)). The results reveal that solvents predicted to be non-resistant (i.e., more likely to dissolve polymers) are 
concentrated in the low-polarity region, whereas solvents predicted to be resistant are distributed in the high-polarity region.

This result aligns with previous reports on the classical “like dissolves like” solubility rule, indicating that the model’s predictions reflect established chemical intuition.

The relationship between solvent polarity and polymer solubility or swelling behavior has been extensively supported by previous studies. 
For example, Toolan et al. reported that in poly(ethylene oxide) (PEO) thin films, the polarity of the solvent significantly affects molecular chain orientation 
and crystallization behavior.\cite{Toolan2016} Lang et al. demonstrated that in non-entangled gels, both the degree of swelling and the residual bond orientation vary markedly 
depending on the strength of polymer--solvent interactions.\cite{Lang2022} In addition, Aoki et al. showed that polymer--solvent compatibility can be predicted 
using machine learning models that incorporate solvent polarity as a key descriptor.\cite{Aoki2023}

\begin{figure*}[t]
    \centering
    \includegraphics[width=\linewidth]{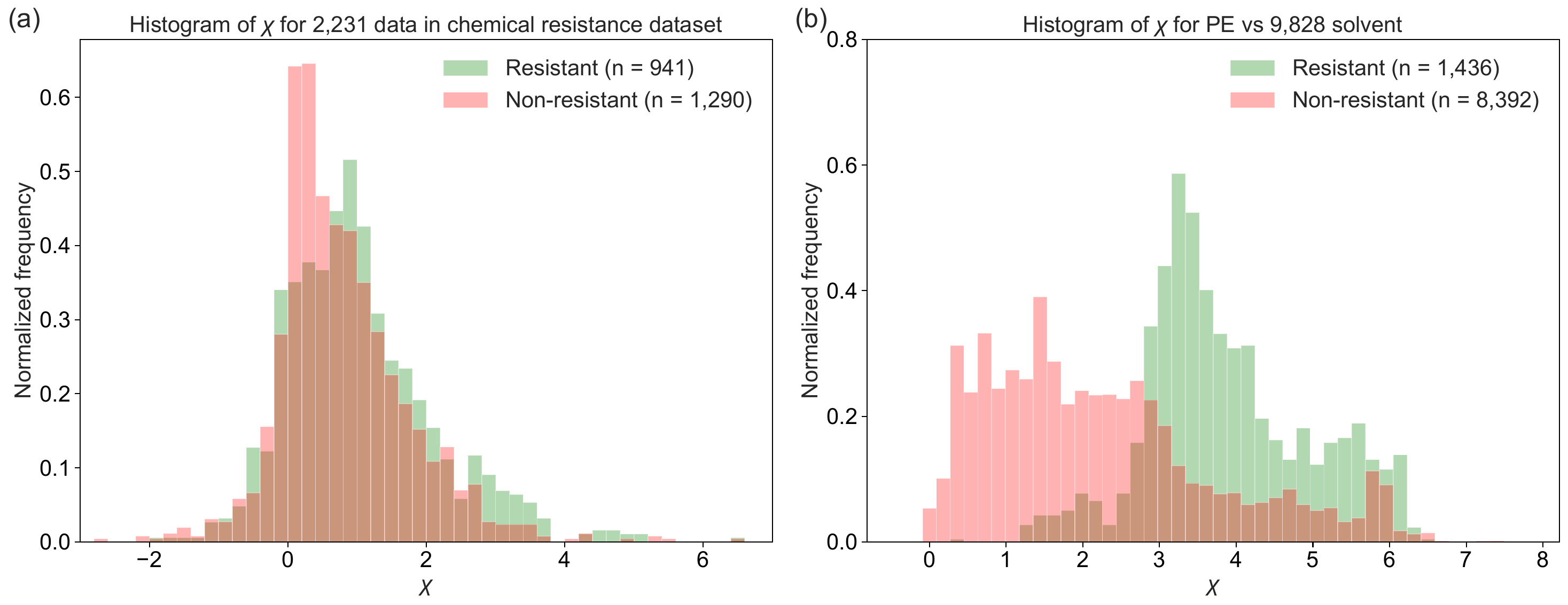}
    \caption{Histograms showing the relationship between the $\chi$ parameter and chemical resistance. (a) Distribution of $\chi$ parameters calculated using the COSMO method (x-axis) and corresponding data density (y-axis) for 2,231 polymer–solvent pairs used in model training. Green and red bars represent resistant and non-resistant cases, respectively. (b) Distribution of predicted $\chi$ parameters for PE with 9,828 solvents, as estimated by the machine learning model. The x-axis indicates predicted $\chi$ values, and the y-axis shows data density. Resistance predictions based on the chemical resistance model are color-coded as green (resistant) and red (non-resistant).}
    \label{fig:solvent_analysis}
\end{figure*}

Furthermore, Navarro et al. demonstrated that solvent polarity enables structural control of block copolymers, 
while Dugas et al. reported a correlation between solvent polarity and membrane permeability.\cite{Navarro2022,Dugas2023} Nezili et al. found that 
highly polar solvents tend to swell and degrade elastomers more readily.\cite{Nezili2023}

This finding supports existing theories, 
supporting the conclusion that polarity is a valid and influential descriptor for predicting chemical resistance.

Figure 8b presents a corresponding analysis for $K_{\text{angle}}$, 
the bond angle force constant, which quantifies the stiffness of the angle formed by three bonded atoms in a molecule. 
A higher $K_{\text{angle}}$ value reflects greater structural rigidity. Solvents predicted to exhibit higher chemical resistance tend to have larger $K_{\text{angle}}$ values, 
indicating that structurally rigid solvents are less capable of penetrating into or swelling polymer chains. 
This trend is consistent with established force field parameter systems, 
where bonds involving highly electronegative atoms---such as oxygen and nitrogen---are known to exhibit higher $K_{\text{angle}}$ values.\cite{Jorgensen1996,Jorge2021}

In summary, the two distinct solvent descriptors---polar (polarity) and $K_{\text{angle}}$ (structural rigidity)---contribute complementarily to chemical resistance prediction. 
The consistency between the trends captured by the model and existing chemical knowledge further validates the reliability of the data-driven predictive framework 
employed in this study.

\subsection*{Solubility parameter}

Figure 9 shows the distribution of Flory--Huggins interaction parameters ($\chi$) for polymer--solvent pairs.
Figure 9a presents the $\chi$ parameter distribution calculated using the COSMO method for the experimental
chemical resistance dataset consisting of 2,231 entries, while Figure 9b shows the results of applying
the model to the combinations of PE with 9,828 solvents.
In both plots, the data are stratified and displayed as histograms according to the presence or absence
of chemical resistance.

In Figure 9a, no clear difference was observed in the $\chi$ parameter distributions between resistant and
non-resistant pairs in the experimental dataset. One likely reason is the diverse nature of the polymers
included in the dataset, which spans a broad range of resistance levels. To clarify the relationship
between $\chi$ and chemical resistance, we recalculated the $\chi$ parameter distributions for each polymer
individually by evaluating all combinations of 27 polymers with the 9,828 solvents and conducted
a stratified analysis. The full distributions for all 27 polymers are provided in Figure S6.

When examining the $\chi$ parameter distributions by resistance label, polymers with weak or intermediate
chemical resistance showed clear separation between resistant and non-resistant cases. In particular,
polymers such as polyethylene (PE), polypropylene (PP), and polymethylpentene (PMP)---all of which exhibit moderate resistance---displayed strong separation
in $\chi$ parameter distributions. In contrast, for highly resistant polymers, such separation was not apparent.

PE, PP, and PMP---all of which showed moderate
resistance---are classified as semicrystalline polymers. In fact, the semicrystalline structure of PE,
comprising both crystalline and amorphous domains, has been well documented in nanoplastics research,\cite{Venezia2025}
and for PP, the coexistence of crystalline and amorphous phases has been shown to influence cavitation
behavior under tensile stress.\cite{Pawlak2008} Similarly, for PMP, gas permeability studies using samples with
different crystallinities have explicitly confirmed its semicrystalline nature.\cite{Puleo1989}

Given the semicrystalline nature of these materials, we further interpret the observed separation in $\chi$
parameter distributions between resistant and non-resistant cases by considering existing models of
polymer dissolution.

Polymer dissolution is generally understood to proceed via a two-step mechanism.\cite{Thomas1982,Peppas1994} In the first step,
the solvent penetrates the amorphous regions of the polymer, swelling the material by entering the spaces
between polymer chains. In the second step, this swelling loosens the structure, allowing crystalline
domains to collapse and the entire polymer to become molecularly solubilized (i.e., dispersed in solution).
Such a two-stage process is also observed in non-Fickian diffusion behaviors, such as Case II diffusion,
which is characterized by solvent ingress and viscoelastic relaxation of the polymer occurring on distinct
time scales.\cite{Thomas1982}

\begin{figure*}[t]
  \centering
  \includegraphics[width=\linewidth]{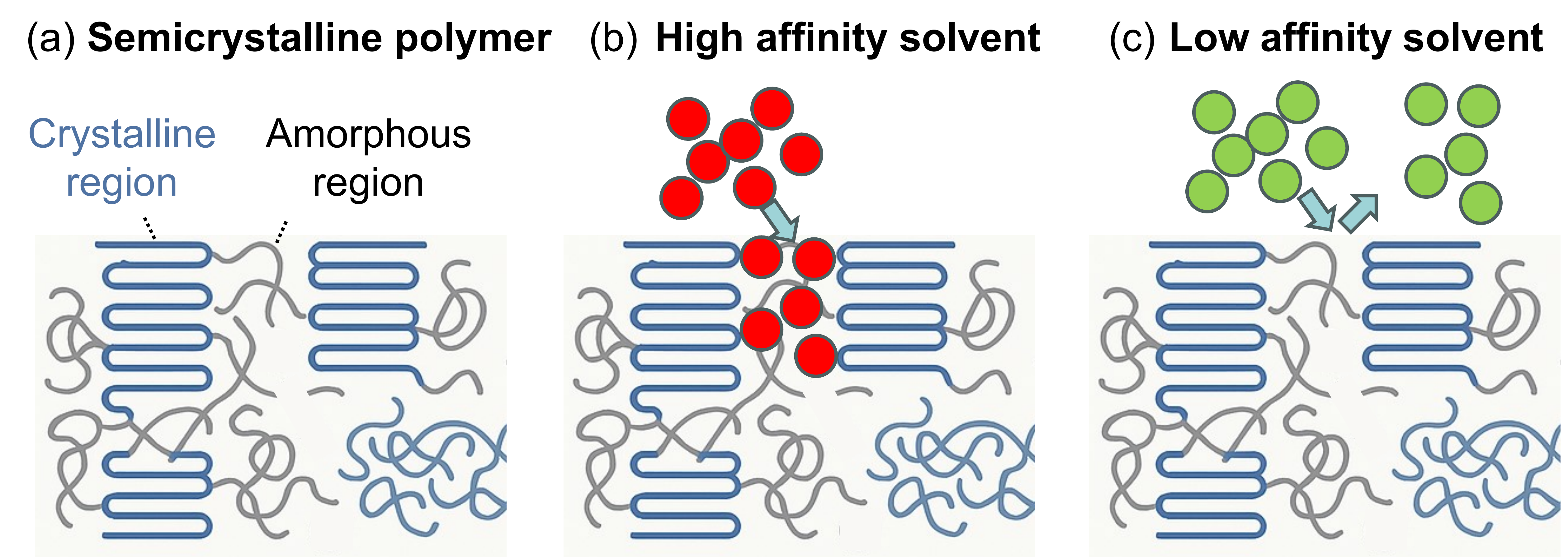}
  \caption{Schematic illustration of solvent penetration depending on the $\chi$ parameter. (a) Model of semicrystalline polymer. (b) When the $\chi$ parameter is small, the solvent penetrates through the amorphous regions. (c) When the $\chi$ parameter is large, solvent penetration is suppressed.}
  \label{fig:fig10}
\end{figure*}

This two-step model is supported by several studies on amorphous and semicrystalline polymers.
For example, both Devotta and Peppas have demonstrated through experimental and theoretical work that
swelling precedes dissolution in amorphous polymers.\cite{Peppas1994,Devotta1994} Similarly, Gardeniers et al. reported that
in semicrystalline polyamides, hydration and swelling of the amorphous phase precede the breakdown of
the crystalline phase during stepwise dissolution.\cite{Gardeniers2022}

In this two-stage model, the Flory--Huggins interaction parameter primarily serves as an index for
quantifying the mixing free energy between the polymer and solvent in the amorphous phase---the first
stage of dissolution. Indeed, lower $\chi$ values have been shown to promote solvent penetration and swelling
within amorphous regions, as demonstrated by Peppas et al. using dissolution--diffusion models based on
polymer--solvent interaction strength,\cite{Peppas1994} and by Ribar et al., who explicitly evaluated the $\chi$ parameter.\cite{Ribar2000}

These findings support the view that the $\chi$ parameter is a useful descriptor for capturing the initial
swelling behavior of semicrystalline polymers.

In Figure 9, the clear separation of the $\chi$ parameter distribution between chemically resistant and
non-resistant semicrystalline polymers suggests a strong correlation between polymer structure and
solvent penetration behavior. As illustrated in Figure 10(a), 
semicrystalline polymers possess a hierarchical structure composed of alternating crystalline layers (lamellae) and 
amorphous regions,\cite{Keller1957,Kanomi2023} and solvent penetration is considered to initiate primarily from the amorphous regions.

Based on this structural characteristic and the results shown in Figure 9, 
it is hypothesized that, as depicted in Figure 10(b),
solvents with low $\chi$ parameters readily penetrate the amorphous regions and
induce swelling by diffusing between polymer chains.
Conversely, as shown in Figure 10(c), solvents with high $\chi$ parameters exhibit limited penetration
into the amorphous regions, resulting in suppressed swelling and dissolution.

For polymers like polymethyl methacrylate (PMMA), which are fully amorphous, dissolution tends to occur almost entirely during the first step, 
as the majority of the structure lacks crystalline domains. Numerous studies have shown that PMMA readily absorbs solvents and moisture, 
leading to rapid swelling.\cite{Peppas1994,Ercken1996} Therefore, in theory, the $\chi$ parameter should be highly sensitive to differences in chemical resistance, 
making it an effective indicator of resistance behavior. However, in the present dataset, many amorphous polymers were generally non-resistant, 
and even solvents with high $\chi$ values were still able to dissolve or swell them. 
This aligns with Peppas' findings that solvent ingress and chain mobilization are directly linked to dissolution behavior in amorphous polymers.\cite{Narasimhan1996} 
It is also important to note that the chemical resistance tests in this study involved prolonged immersion over several days. 
Thus, even solvents that may not cause significant short-term swelling might still be labeled as “non-resistant” after extended exposure. 
Lyu et al. have theoretically shown that even when initial swelling is not observed, polymer chains in amorphous materials can relax and swell over time.\cite{Lyu2024}

In summary, although the $\chi$ parameter effectively describes the thermodynamic interactions in the first stage of dissolution, 
it may be less informative for predicting chemical resistance in amorphous polymers, 
as the distinction in $\chi$ distributions between resistant and non-resistant cases was not evident.

As for highly resistant crystalline polymers such as PTFE, 
the lack of separation in the $\chi$ parameter distributions may be due to the overwhelming number of cases being classified as resistant. 
This limits the variation in $\chi$ values in the model outputs. 
PTFE is known to exhibit a high degree of crystallinity; in fact, differential scanning calorimetry (DSC) measurements on unsintered PTFE powders have shown crystallinity above 90\%, 
and even after melting and sintering, crystallinity levels remain high, typically between 40\% and 80\%.\cite{Brown2005} 
In such highly crystalline polymers, the amorphous regions are extremely limited, 
making solvent penetration into those regions physically difficult. 
As a result, thermodynamic interaction differences expressed by the $\chi$ parameter may not manifest in actual chemical resistance behavior.

Taken together, these findings suggest that the $\chi$ parameter is particularly useful for systems where solvent--polymer interactions in the amorphous phase dominate, 
such as semicrystalline polymers that may eventually progress to the second stage of dissolution over sufficiently long timescales. 
In contrast, for highly crystalline polymers, the $\chi$ parameter may not capture observable differences in chemical resistance.

In conclusion, analyzing the $\chi$ parameter---a chemically meaningful quantity---through the lens of machine learning predictions 
enabled a discussion consistent with established polymer dissolution theories. 
This confirms that data-driven approaches are effective tools for interpreting and predicting chemical resistance.

\vspace{2em}

\section*{Conclusions}

In this study, we conducted a systematic, data-driven analysis to identify key factors influencing the chemical resistance of polymers.
Using chemical resistance data collected from the literature, we constructed predictive models incorporating the FFKM descriptors, MD-derived polymer properties, and Flory--Huggins interaction parameters as explanatory variables for both polymers and solvents.
The resulting models achieved high predictive performance, with average ROC-AUC values of 0.85 and 0.91 for polymer-level and solvent cluster-level cross-validation, respectively, demonstrating the ability to accurately classify chemical resistance.

Feature analysis revealed that polymer crystallinity is a critical factor in predicting chemical resistance.
Additionally, a positive correlation was observed between MD-calculated polymer density and predicted resistance probability, indicating that denser polymers tend to exhibit higher chemical resistance.
On the solvent side, highly polar solvents were more likely to be predicted as resistant, while non-polar solvents tended to be associated with higher solubility of polymers---consistent with the classical ``like dissolves like'' rule.

With regard to the Flory--Huggins interaction parameter ($\chi$), particularly in semicrystalline polymers, clear separation was observed in the $\chi$ distributions between resistant and non-resistant cases.
This finding aligns with the conventional two-step dissolution model, in which solvents first penetrate the amorphous regions, followed by the breakdown of crystalline domains.
The $\chi$ parameter thus serves as a meaningful indicator of the initial swelling phase in such systems.

Overall, the results demonstrate that even with relatively simple explanatory variables and models, it is possible to predict chemical resistance and perform factor analysis in a manner consistent with established chemical understanding.
This study highlights the potential of data-driven approaches as a powerful strategy for predicting and designing chemical resistance in polymers, with promising implications for future materials design and screening applications.

\vspace{3em}

\section*{Acknowledgements}

We thank the members of SCREEN Semiconductor Solutions, Inc., 
for their fruitful discussions and for reviewing the manuscript. 
This research was partly supported by the Ministry of Education, Culture, Sports, Science and Technology (MEXT) 
through the "Program for Promoting Researches on the Supercomputer Fugaku" (JPMXP1020200314) 
and the Japan Science and Technology Agency (JST) (JPMJCR2332). 
Computational resources were provided by Fugaku at the RIKEN Center for Computational Science, 
Kobe, Japan (hp210264, hp220179, hp230190, hp240216, and hp250235).

\bibliography{lib}

\begin{thebibliography}{66}%
\makeatletter
\providecommand \@ifxundefined [1]{%
 \@ifx{#1\undefined}
}%
\providecommand \@ifnum [1]{%
 \ifnum #1\expandafter \@firstoftwo
 \else \expandafter \@secondoftwo
 \fi
}%
\providecommand \@ifx [1]{%
 \ifx #1\expandafter \@firstoftwo
 \else \expandafter \@secondoftwo
 \fi
}%
\providecommand \natexlab [1]{#1}%
\providecommand \enquote  [1]{``#1''}%
\providecommand \bibnamefont  [1]{#1}%
\providecommand \bibfnamefont [1]{#1}%
\providecommand \citenamefont [1]{#1}%
\providecommand \href@noop [0]{\@secondoftwo}%
\providecommand \href [0]{\begingroup \@sanitize@url \@href}%
\providecommand \@href[1]{\@@startlink{#1}\@@href}%
\providecommand \@@href[1]{\endgroup#1\@@endlink}%
\providecommand \@sanitize@url [0]{\catcode `\\12\catcode `\$12\catcode
  `\&12\catcode `\#12\catcode `\^12\catcode `\_12\catcode `\%12\relax}%
\providecommand \@@startlink[1]{}%
\providecommand \@@endlink[0]{}%
\providecommand \url  [0]{\begingroup\@sanitize@url \@url }%
\providecommand \@url [1]{\endgroup\@href {#1}{\urlprefix }}%
\providecommand \urlprefix  [0]{URL }%
\providecommand \Eprint [0]{\href }%
\providecommand \doibase [0]{https://doi.org/}%
\providecommand \selectlanguage [0]{\@gobble}%
\providecommand \bibinfo  [0]{\@secondoftwo}%
\providecommand \bibfield  [0]{\@secondoftwo}%
\providecommand \translation [1]{[#1]}%
\providecommand \BibitemOpen [0]{}%
\providecommand \bibitemStop [0]{}%
\providecommand \bibitemNoStop [0]{.\EOS\space}%
\providecommand \EOS [0]{\spacefactor3000\relax}%
\providecommand \BibitemShut  [1]{\csname bibitem#1\endcsname}%
\let\auto@bib@innerbib\@empty
\bibitem [{\citenamefont {Hansen}(1967)}]{Hansen1967}%
  \BibitemOpen
  \bibfield  {author} {\bibinfo {author} {\bibfnamefont {C.}~\bibnamefont
  {Hansen}},\ }\href@noop {} {\emph {\bibinfo {title} {The {{Three Dimensional
  Solubility Parameter}} and {{Solvent Diffusion Coefficient}} and {{Their
  Importance}} in {{Surface Coating Formulation}}}}}\ (\bibinfo  {publisher}
  {Danish Technical Press},\ \bibinfo {address} {Copenhagen, Denmark},\
  \bibinfo {year} {1967})\BibitemShut {NoStop}%
\bibitem [{\citenamefont {Hansen}(2007)}]{Hansen2007}%
  \BibitemOpen
  \bibfield  {author} {\bibinfo {author} {\bibfnamefont {C.}~\bibnamefont
  {Hansen}},\ }\href@noop {} {\emph {\bibinfo {title} {Hansen {{Solubility
  Parameter}}: {{A User}}'s {{Handbook}}}}},\ \bibinfo {edition} {2nd}\ ed.\
  (\bibinfo  {publisher} {CRC Press},\ \bibinfo {year} {2007})\BibitemShut
  {NoStop}%
\bibitem [{\citenamefont {Zhou}\ \emph {et~al.}(2023)\citenamefont {Zhou},
  \citenamefont {Yu}, \citenamefont {{S{\'a}nchez-Rivera}}, \citenamefont
  {Huber},\ and\ \citenamefont {Lehn}}]{Zhou2023}%
  \BibitemOpen
  \bibfield  {author} {\bibinfo {author} {\bibfnamefont {P.}~\bibnamefont
  {Zhou}}, \bibinfo {author} {\bibfnamefont {J.}~\bibnamefont {Yu}}, \bibinfo
  {author} {\bibfnamefont {K.~L.}\ \bibnamefont {{S{\'a}nchez-Rivera}}},
  \bibinfo {author} {\bibfnamefont {G.~W.}\ \bibnamefont {Huber}},\ and\
  \bibinfo {author} {\bibfnamefont {R.~C.~V.}\ \bibnamefont {Lehn}},\
  }\bibfield  {title} {\bibinfo {title} {Large-scale computational polymer
  solubility predictions and applications to dissolution-based plastic
  recycling},\ }\href {https://doi.org/10.1039/D3GC00404J} {\bibfield
  {journal} {\bibinfo  {journal} {Green Chemistry}\ }\textbf {\bibinfo {volume}
  {25}},\ \bibinfo {pages} {4402} (\bibinfo {year} {2023})}\BibitemShut
  {NoStop}%
\bibitem [{\citenamefont {Peppas}\ \emph {et~al.}(2006)\citenamefont {Peppas},
  \citenamefont {Hilt}, \citenamefont {Khademhosseini},\ and\ \citenamefont
  {Langer}}]{Peppas2006}%
  \BibitemOpen
  \bibfield  {author} {\bibinfo {author} {\bibfnamefont {N.~A.}\ \bibnamefont
  {Peppas}}, \bibinfo {author} {\bibfnamefont {J.~Z.}\ \bibnamefont {Hilt}},
  \bibinfo {author} {\bibfnamefont {A.}~\bibnamefont {Khademhosseini}},\ and\
  \bibinfo {author} {\bibfnamefont {R.}~\bibnamefont {Langer}},\ }\bibfield
  {title} {\bibinfo {title} {Hydrogels in {{Biology}} and {{Medicine}}: {{From
  Molecular Principles}} to {{Bionanotechnology}}},\ }\href
  {https://doi.org/10.1002/adma.200501612} {\bibfield  {journal} {\bibinfo
  {journal} {Advanced Materials}\ }\textbf {\bibinfo {volume} {18}},\ \bibinfo
  {pages} {1345} (\bibinfo {year} {2006})}\BibitemShut {NoStop}%
\bibitem [{\citenamefont {Zhou}\ \emph {et~al.}(2018)\citenamefont {Zhou},
  \citenamefont {Zhang}, \citenamefont {Yang},\ and\ \citenamefont
  {Wu}}]{Zhou2018}%
  \BibitemOpen
  \bibfield  {author} {\bibinfo {author} {\bibfnamefont {Q.}~\bibnamefont
  {Zhou}}, \bibinfo {author} {\bibfnamefont {L.}~\bibnamefont {Zhang}},
  \bibinfo {author} {\bibfnamefont {T.}~\bibnamefont {Yang}},\ and\ \bibinfo
  {author} {\bibfnamefont {H.}~\bibnamefont {Wu}},\ }\bibfield  {title}
  {\bibinfo {title} {Stimuli-responsive polymeric micelles for drug delivery
  and cancer therapy},\ }\href {https://doi.org/10.2147/IJN.S158696} {\bibfield
   {journal} {\bibinfo  {journal} {International Journal of Nanomedicine}\
  }\textbf {\bibinfo {volume} {13}},\ \bibinfo {pages} {2921} (\bibinfo {year}
  {2018})}\BibitemShut {NoStop}%
\bibitem [{\citenamefont {Blum}\ \emph {et~al.}(2015)\citenamefont {Blum},
  \citenamefont {Balan}, \citenamefont {Scheringer}, \citenamefont {Trier},
  \citenamefont {Goldenman}, \citenamefont {Cousins}, \citenamefont {Diamond},
  \citenamefont {Fletcher}, \citenamefont {Higgins}, \citenamefont {Lindeman},
  \citenamefont {Peaslee}, \citenamefont {{de Voogt}}, \citenamefont {Wang},\
  and\ \citenamefont {Weber}}]{Blum2015}%
  \BibitemOpen
  \bibfield  {author} {\bibinfo {author} {\bibfnamefont {A.}~\bibnamefont
  {Blum}}, \bibinfo {author} {\bibfnamefont {S.~A.}\ \bibnamefont {Balan}},
  \bibinfo {author} {\bibfnamefont {M.}~\bibnamefont {Scheringer}}, \bibinfo
  {author} {\bibfnamefont {X.}~\bibnamefont {Trier}}, \bibinfo {author}
  {\bibfnamefont {G.}~\bibnamefont {Goldenman}}, \bibinfo {author}
  {\bibfnamefont {I.~T.}\ \bibnamefont {Cousins}}, \bibinfo {author}
  {\bibfnamefont {M.}~\bibnamefont {Diamond}}, \bibinfo {author} {\bibfnamefont
  {T.}~\bibnamefont {Fletcher}}, \bibinfo {author} {\bibfnamefont
  {C.}~\bibnamefont {Higgins}}, \bibinfo {author} {\bibfnamefont {A.~E.}\
  \bibnamefont {Lindeman}}, \bibinfo {author} {\bibfnamefont {G.}~\bibnamefont
  {Peaslee}}, \bibinfo {author} {\bibfnamefont {P.}~\bibnamefont {{de Voogt}}},
  \bibinfo {author} {\bibfnamefont {Z.}~\bibnamefont {Wang}},\ and\ \bibinfo
  {author} {\bibfnamefont {R.}~\bibnamefont {Weber}},\ }\bibfield  {title}
  {\bibinfo {title} {The {{Madrid Statement}} on {{Poly-}} and {{Perfluoroalkyl
  Substances}} ({{PFASs}})},\ }\href {https://doi.org/10.1289/ehp.1509934}
  {\bibfield  {journal} {\bibinfo  {journal} {Environmental Health
  Perspectives}\ }\textbf {\bibinfo {volume} {123}},\ \bibinfo {pages} {A107}
  (\bibinfo {year} {2015})}\BibitemShut {NoStop}%
\bibitem [{\citenamefont {Brown}\ \emph {et~al.}(2020)\citenamefont {Brown},
  \citenamefont {Conder}, \citenamefont {Arblaster},\ and\ \citenamefont
  {Higgins}}]{Brown2020}%
  \BibitemOpen
  \bibfield  {author} {\bibinfo {author} {\bibfnamefont {J.~B.}\ \bibnamefont
  {Brown}}, \bibinfo {author} {\bibfnamefont {J.~M.}\ \bibnamefont {Conder}},
  \bibinfo {author} {\bibfnamefont {J.~A.}\ \bibnamefont {Arblaster}},\ and\
  \bibinfo {author} {\bibfnamefont {C.~P.}\ \bibnamefont {Higgins}},\
  }\bibfield  {title} {\bibinfo {title} {Assessing {{Human Health Risks}} from
  {{Per-}} and {{Polyfluoroalkyl Substance}} ({{PFAS}})-{{Impacted Vegetable
  Consumption}}: {{A Tiered Modeling Approach}}},\ }\href
  {https://doi.org/10.1021/acs.est.0c03411} {\bibfield  {journal} {\bibinfo
  {journal} {Environmental Science \& Technology}\ }\textbf {\bibinfo {volume}
  {54}},\ \bibinfo {pages} {15202} (\bibinfo {year} {2020})}\BibitemShut
  {NoStop}%
\bibitem [{\citenamefont {Cousins}\ \emph {et~al.}(2019)\citenamefont
  {Cousins}, \citenamefont {Goldenman}, \citenamefont {Herzke}, \citenamefont
  {Lohmann}, \citenamefont {Miller}, \citenamefont {Ng}, \citenamefont
  {Patton}, \citenamefont {Scheringer}, \citenamefont {Trier}, \citenamefont
  {Vierke}, \citenamefont {Wang},\ and\ \citenamefont {DeWitt}}]{Cousins2019}%
  \BibitemOpen
  \bibfield  {author} {\bibinfo {author} {\bibfnamefont {I.~T.}\ \bibnamefont
  {Cousins}}, \bibinfo {author} {\bibfnamefont {G.}~\bibnamefont {Goldenman}},
  \bibinfo {author} {\bibfnamefont {D.}~\bibnamefont {Herzke}}, \bibinfo
  {author} {\bibfnamefont {R.}~\bibnamefont {Lohmann}}, \bibinfo {author}
  {\bibfnamefont {M.}~\bibnamefont {Miller}}, \bibinfo {author} {\bibfnamefont
  {C.~A.}\ \bibnamefont {Ng}}, \bibinfo {author} {\bibfnamefont
  {S.}~\bibnamefont {Patton}}, \bibinfo {author} {\bibfnamefont
  {M.}~\bibnamefont {Scheringer}}, \bibinfo {author} {\bibfnamefont
  {X.}~\bibnamefont {Trier}}, \bibinfo {author} {\bibfnamefont
  {L.}~\bibnamefont {Vierke}}, \bibinfo {author} {\bibfnamefont
  {Z.}~\bibnamefont {Wang}},\ and\ \bibinfo {author} {\bibfnamefont {J.~C.}\
  \bibnamefont {DeWitt}},\ }\bibfield  {title} {\bibinfo {title} {The concept
  of essential use for determining when uses of {{PFASs}} can be phased out},\
  }\href {https://doi.org/10.1039/C9EM00163H} {\bibfield  {journal} {\bibinfo
  {journal} {Environmental Science: Processes \& Impacts}\ }\textbf {\bibinfo
  {volume} {21}},\ \bibinfo {pages} {1803} (\bibinfo {year}
  {2019})}\BibitemShut {NoStop}%
\bibitem [{\citenamefont {Hildebrand}\ and\ \citenamefont
  {Scott}(1950)}]{Hildebrand1950}%
  \BibitemOpen
  \bibfield  {author} {\bibinfo {author} {\bibfnamefont {J.~H.}\ \bibnamefont
  {Hildebrand}}\ and\ \bibinfo {author} {\bibfnamefont {R.~L.}\ \bibnamefont
  {Scott}},\ }\href@noop {} {\emph {\bibinfo {title} {The {{Solubility}} of
  {{Nonelectrolytes}}}}},\ \bibinfo {edition} {3rd}\ ed.\ (\bibinfo
  {publisher} {Reinhold Publishing Corp.},\ \bibinfo {address} {New York},\
  \bibinfo {year} {1950})\BibitemShut {NoStop}%
\bibitem [{\citenamefont {Flory}(1942)}]{Flory1942}%
  \BibitemOpen
  \bibfield  {author} {\bibinfo {author} {\bibfnamefont {P.~J.}\ \bibnamefont
  {Flory}},\ }\bibfield  {title} {\bibinfo {title} {Thermodynamics of {{High
  Polymer Solutions}}},\ }\href {https://doi.org/10.1063/1.1723621} {\bibfield
  {journal} {\bibinfo  {journal} {The Journal of Chemical Physics}\ }\textbf
  {\bibinfo {volume} {10}},\ \bibinfo {pages} {51} (\bibinfo {year}
  {1942})}\BibitemShut {NoStop}%
\bibitem [{\citenamefont {Huggins}(1941)}]{Huggins1941}%
  \BibitemOpen
  \bibfield  {author} {\bibinfo {author} {\bibfnamefont {M.~L.}\ \bibnamefont
  {Huggins}},\ }\bibfield  {title} {\bibinfo {title} {Solutions of {{Long Chain
  Compounds}}},\ }\href {https://doi.org/10.1063/1.1750930} {\bibfield
  {journal} {\bibinfo  {journal} {The Journal of Chemical Physics}\ }\textbf
  {\bibinfo {volume} {9}},\ \bibinfo {pages} {440} (\bibinfo {year}
  {1941})}\BibitemShut {NoStop}%
\bibitem [{\citenamefont {Kim}\ \emph {et~al.}(2018)\citenamefont {Kim},
  \citenamefont {Chandrasekaran}, \citenamefont {Huan}, \citenamefont {Das},\
  and\ \citenamefont {Ramprasad}}]{Kim2018}%
  \BibitemOpen
  \bibfield  {author} {\bibinfo {author} {\bibfnamefont {C.}~\bibnamefont
  {Kim}}, \bibinfo {author} {\bibfnamefont {A.}~\bibnamefont {Chandrasekaran}},
  \bibinfo {author} {\bibfnamefont {T.~D.}\ \bibnamefont {Huan}}, \bibinfo
  {author} {\bibfnamefont {D.}~\bibnamefont {Das}},\ and\ \bibinfo {author}
  {\bibfnamefont {R.}~\bibnamefont {Ramprasad}},\ }\bibfield  {title} {\bibinfo
  {title} {Polymer {{Genome}}: {{A Data-Powered Polymer Informatics Platform}}
  for {{Property Predictions}}},\ }\href
  {https://doi.org/10.1021/acs.jpcc.8b02913} {\bibfield  {journal} {\bibinfo
  {journal} {The Journal of Physical Chemistry C}\ }\textbf {\bibinfo {volume}
  {122}},\ \bibinfo {pages} {17575} (\bibinfo {year} {2018})}\BibitemShut
  {NoStop}%
\bibitem [{\citenamefont {Xu}\ and\ \citenamefont {Jiang}(2020)}]{Xu2020}%
  \BibitemOpen
  \bibfield  {author} {\bibinfo {author} {\bibfnamefont {Q.}~\bibnamefont
  {Xu}}\ and\ \bibinfo {author} {\bibfnamefont {J.}~\bibnamefont {Jiang}},\
  }\bibfield  {title} {\bibinfo {title} {Machine {{Learning}} for {{Polymer
  Swelling}} in {{Liquids}}},\ }\href {https://doi.org/10.1021/acsapm.0c00586}
  {\bibfield  {journal} {\bibinfo  {journal} {ACS Applied Polymer Materials}\
  }\textbf {\bibinfo {volume} {2}},\ \bibinfo {pages} {3576} (\bibinfo {year}
  {2020})}\BibitemShut {NoStop}%
\bibitem [{\citenamefont {Aoki}\ \emph {et~al.}(2023)\citenamefont {Aoki},
  \citenamefont {Wu}, \citenamefont {Tsurimoto}, \citenamefont {Hayashi},
  \citenamefont {Minami}, \citenamefont {Tadamichi}, \citenamefont
  {Shiratori},\ and\ \citenamefont {Yoshida}}]{Aoki2023}%
  \BibitemOpen
  \bibfield  {author} {\bibinfo {author} {\bibfnamefont {Y.}~\bibnamefont
  {Aoki}}, \bibinfo {author} {\bibfnamefont {S.}~\bibnamefont {Wu}}, \bibinfo
  {author} {\bibfnamefont {T.}~\bibnamefont {Tsurimoto}}, \bibinfo {author}
  {\bibfnamefont {Y.}~\bibnamefont {Hayashi}}, \bibinfo {author} {\bibfnamefont
  {S.}~\bibnamefont {Minami}}, \bibinfo {author} {\bibfnamefont
  {O.}~\bibnamefont {Tadamichi}}, \bibinfo {author} {\bibfnamefont
  {K.}~\bibnamefont {Shiratori}},\ and\ \bibinfo {author} {\bibfnamefont
  {R.}~\bibnamefont {Yoshida}},\ }\bibfield  {title} {\bibinfo {title}
  {Multitask {{Machine Learning}} to {{Predict Polymer}}--{{Solvent Miscibility
  Using Flory}}--{{Huggins Interaction Parameters}}},\ }\href
  {https://doi.org/10.1021/acs.macromol.2c02600} {\bibfield  {journal}
  {\bibinfo  {journal} {Macromolecules}\ }\textbf {\bibinfo {volume} {56}},\
  \bibinfo {pages} {5446} (\bibinfo {year} {2023})}\BibitemShut {NoStop}%
\bibitem [{\citenamefont {Yu}\ \emph {et~al.}(2023)\citenamefont {Yu},
  \citenamefont {Zhang}, \citenamefont {Cheng}, \citenamefont {Yang},
  \citenamefont {She}, \citenamefont {Liu}, \citenamefont {Su},\ and\
  \citenamefont {Su}}]{Yu2023}%
  \BibitemOpen
  \bibfield  {author} {\bibinfo {author} {\bibfnamefont {J.}~\bibnamefont
  {Yu}}, \bibinfo {author} {\bibfnamefont {C.}~\bibnamefont {Zhang}}, \bibinfo
  {author} {\bibfnamefont {Y.}~\bibnamefont {Cheng}}, \bibinfo {author}
  {\bibfnamefont {Y.-F.}\ \bibnamefont {Yang}}, \bibinfo {author}
  {\bibfnamefont {Y.-B.}\ \bibnamefont {She}}, \bibinfo {author} {\bibfnamefont
  {F.}~\bibnamefont {Liu}}, \bibinfo {author} {\bibfnamefont {W.}~\bibnamefont
  {Su}},\ and\ \bibinfo {author} {\bibfnamefont {A.}~\bibnamefont {Su}},\
  }\bibfield  {title} {\bibinfo {title} {{{SolvBERT}} for solvation free energy
  and solubility prediction: A demonstration of an {{NLP}} model for predicting
  the properties of molecular complexes},\ }\href
  {https://doi.org/10.1039/D2DD00107A} {\bibfield  {journal} {\bibinfo
  {journal} {Digital Discovery}\ }\textbf {\bibinfo {volume} {2}},\ \bibinfo
  {pages} {409} (\bibinfo {year} {2023})}\BibitemShut {NoStop}%
\bibitem [{\citenamefont {Kuenneth}\ and\ \citenamefont
  {Ramprasad}(2023)}]{Kuenneth2023}%
  \BibitemOpen
  \bibfield  {author} {\bibinfo {author} {\bibfnamefont {C.}~\bibnamefont
  {Kuenneth}}\ and\ \bibinfo {author} {\bibfnamefont {R.}~\bibnamefont
  {Ramprasad}},\ }\bibfield  {title} {\bibinfo {title} {{{polyBERT}}: A
  chemical language model to enable fully machine-driven ultrafast polymer
  informatics},\ }\href {https://doi.org/10.1038/s41467-023-39868-6} {\bibfield
   {journal} {\bibinfo  {journal} {Nature Communications}\ }\textbf {\bibinfo
  {volume} {14}},\ \bibinfo {pages} {4099} (\bibinfo {year}
  {2023})}\BibitemShut {NoStop}%
\bibitem [{\citenamefont {Agarwal}\ \emph {et~al.}(2025)\citenamefont
  {Agarwal}, \citenamefont {Mahmood},\ and\ \citenamefont
  {Ramprasad}}]{Agarwal2025}%
  \BibitemOpen
  \bibfield  {author} {\bibinfo {author} {\bibfnamefont {S.}~\bibnamefont
  {Agarwal}}, \bibinfo {author} {\bibfnamefont {A.}~\bibnamefont {Mahmood}},\
  and\ \bibinfo {author} {\bibfnamefont {R.}~\bibnamefont {Ramprasad}},\
  }\bibfield  {title} {\bibinfo {title} {Polymer {{Solubility Prediction Using
  Large Language Models}}},\ }\href
  {https://doi.org/10.1021/acsmaterialslett.5c00054} {\bibfield  {journal}
  {\bibinfo  {journal} {ACS Materials Letters}\ ,\ \bibinfo {pages} {2017}}
  (\bibinfo {year} {2025})}\BibitemShut {NoStop}%
\bibitem [{\citenamefont {{Kayo Corporation}}()}]{Kayo}%
  \BibitemOpen
  \bibfield  {author} {\bibinfo {author} {\bibnamefont {{Kayo Corporation}}},\
  }\href@noop {} {\bibinfo {title} {Chemical {{Resistance}} of
  {{Plastics}}}}\BibitemShut {NoStop}%
\bibitem [{\citenamefont {{ASOH Co., Ltd.}}()}]{ASOH}%
  \BibitemOpen
  \bibfield  {author} {\bibinfo {author} {\bibnamefont {{ASOH Co., Ltd.}}},\
  }\href@noop {} {\bibinfo {title} {Chemical {{Resistance Chart}}}}\BibitemShut
  {NoStop}%
\bibitem [{\citenamefont {{PISCO Corporation}}()}]{PISCOCorporation}%
  \BibitemOpen
  \bibfield  {author} {\bibinfo {author} {\bibnamefont {{PISCO Corporation}}},\
  }\href@noop {} {\bibinfo {title} {Chemical {{Resistance Data}}}}\BibitemShut
  {NoStop}%
\bibitem [{\citenamefont {{VICI AG International}}()}]{VICIAGInternational}%
  \BibitemOpen
  \bibfield  {author} {\bibinfo {author} {\bibnamefont {{VICI AG
  International}}},\ }\href@noop {} {\bibinfo {title} {Chemical resistance
  {{PEEK}} and other polymers}}\BibitemShut {NoStop}%
\bibitem [{\citenamefont {{Professional Plastics}}()}]{ProfessionalPlastics}%
  \BibitemOpen
  \bibfield  {author} {\bibinfo {author} {\bibnamefont {{Professional
  Plastics}}},\ }\href@noop {} {\bibinfo {title} {Chemical {{Resistance
  PCTFE}}}}\BibitemShut {NoStop}%
\bibitem [{\citenamefont {{Thermo Fisher
  Scientific}}()}]{ThermoFisherScientific}%
  \BibitemOpen
  \bibfield  {author} {\bibinfo {author} {\bibnamefont {{Thermo Fisher
  Scientific}}},\ }\href@noop {} {\bibinfo {title} {Labware {{Chemical
  Resistance Table}}}}\BibitemShut {NoStop}%
\bibitem [{\citenamefont {Hayashi}\ \emph {et~al.}(2022)\citenamefont
  {Hayashi}, \citenamefont {Shiomi}, \citenamefont {Morikawa},\ and\
  \citenamefont {Yoshida}}]{Hayashi2022}%
  \BibitemOpen
  \bibfield  {author} {\bibinfo {author} {\bibfnamefont {Y.}~\bibnamefont
  {Hayashi}}, \bibinfo {author} {\bibfnamefont {J.}~\bibnamefont {Shiomi}},
  \bibinfo {author} {\bibfnamefont {J.}~\bibnamefont {Morikawa}},\ and\
  \bibinfo {author} {\bibfnamefont {R.}~\bibnamefont {Yoshida}},\ }\bibfield
  {title} {\bibinfo {title} {{{RadonPy}}: Automated physical property
  calculation using all-atom classical molecular dynamics simulations for
  polymer informatics},\ }\href {https://doi.org/10.1038/s41524-022-00906-4}
  {\bibfield  {journal} {\bibinfo  {journal} {npj Computational Materials}\
  }\textbf {\bibinfo {volume} {8}},\ \bibinfo {pages} {1} (\bibinfo {year}
  {2022})}\BibitemShut {NoStop}%
\bibitem [{HSP()}]{HSP}%
  \BibitemOpen
  \href@noop {} {\bibinfo {title} {Hansen {{Solubility Parameters}} in
  {{Practice}}}}\BibitemShut {NoStop}%
\bibitem [{\citenamefont {Kusaba}\ \emph {et~al.}(2023)\citenamefont {Kusaba},
  \citenamefont {Hayashi}, \citenamefont {Liu}, \citenamefont {Wakiuchi},\ and\
  \citenamefont {Yoshida}}]{Kusaba2023}%
  \BibitemOpen
  \bibfield  {author} {\bibinfo {author} {\bibfnamefont {M.}~\bibnamefont
  {Kusaba}}, \bibinfo {author} {\bibfnamefont {Y.}~\bibnamefont {Hayashi}},
  \bibinfo {author} {\bibfnamefont {C.}~\bibnamefont {Liu}}, \bibinfo {author}
  {\bibfnamefont {A.}~\bibnamefont {Wakiuchi}},\ and\ \bibinfo {author}
  {\bibfnamefont {R.}~\bibnamefont {Yoshida}},\ }\bibfield  {title} {\bibinfo
  {title} {Representation of materials by kernel mean embedding},\ }\href
  {https://doi.org/10.1103/PhysRevB.108.134107} {\bibfield  {journal} {\bibinfo
   {journal} {Physical Review B}\ }\textbf {\bibinfo {volume} {108}},\ \bibinfo
  {pages} {134107} (\bibinfo {year} {2023})}\BibitemShut {NoStop}%
\bibitem [{\citenamefont {Wang}\ \emph {et~al.}(2004)\citenamefont {Wang},
  \citenamefont {Wolf}, \citenamefont {Caldwell}, \citenamefont {Kollman},\
  and\ \citenamefont {Case}}]{Wang2004}%
  \BibitemOpen
  \bibfield  {author} {\bibinfo {author} {\bibfnamefont {J.}~\bibnamefont
  {Wang}}, \bibinfo {author} {\bibfnamefont {R.~M.}\ \bibnamefont {Wolf}},
  \bibinfo {author} {\bibfnamefont {J.~W.}\ \bibnamefont {Caldwell}}, \bibinfo
  {author} {\bibfnamefont {P.~A.}\ \bibnamefont {Kollman}},\ and\ \bibinfo
  {author} {\bibfnamefont {D.~A.}\ \bibnamefont {Case}},\ }\bibfield  {title}
  {\bibinfo {title} {Development and testing of a general amber force field},\
  }\href {https://doi.org/10.1002/jcc.20035} {\bibfield  {journal} {\bibinfo
  {journal} {Journal of Computational Chemistry}\ }\textbf {\bibinfo {volume}
  {25}},\ \bibinfo {pages} {1157} (\bibinfo {year} {2004})}\BibitemShut
  {NoStop}%
\bibitem [{\citenamefont {Muandet}\ \emph {et~al.}(2017)\citenamefont
  {Muandet}, \citenamefont {Fukumizu}, \citenamefont {Sriperumbudur},\ and\
  \citenamefont {Sch{\"o}lkopf}}]{Muandet2017}%
  \BibitemOpen
  \bibfield  {author} {\bibinfo {author} {\bibfnamefont {K.}~\bibnamefont
  {Muandet}}, \bibinfo {author} {\bibfnamefont {K.}~\bibnamefont {Fukumizu}},
  \bibinfo {author} {\bibfnamefont {B.}~\bibnamefont {Sriperumbudur}},\ and\
  \bibinfo {author} {\bibfnamefont {B.}~\bibnamefont {Sch{\"o}lkopf}},\
  }\bibfield  {title} {\bibinfo {title} {Kernel {{Mean Embedding}} of
  {{Distributions}}: {{A Review}} and {{Beyond}}},\ }\href
  {https://doi.org/10.1561/2200000060} {\bibfield  {journal} {\bibinfo
  {journal} {Foundations and Trends{\textregistered} in Machine Learning}\
  }\textbf {\bibinfo {volume} {10}},\ \bibinfo {pages} {1} (\bibinfo {year}
  {2017})}\BibitemShut {NoStop}%
\bibitem [{Rad(2025)}]{RadonPyGithub}%
  \BibitemOpen
  \href@noop {} {\bibinfo {title} {{{RadonPy Github}} site}},\ \bibinfo
  {howpublished} {RadonPy project} (\bibinfo {year} {2025})\BibitemShut
  {NoStop}%
\bibitem [{TUR()}]{TURBOMOLE}%
  \BibitemOpen
  \href@noop {} {\bibinfo {title} {{{TURBOMOLE}} {\textbar} {{Program Package}}
  for {{Electronic Structure Calculations}}}},\ \bibinfo {howpublished}
  {https://www.turbomole.org/}\BibitemShut {NoStop}%
\bibitem [{\citenamefont {{COSMOlogic GmbH \& Co. KG,}}()}]{COSMOthermo}%
  \BibitemOpen
  \bibfield  {author} {\bibinfo {author} {\bibnamefont {{COSMOlogic GmbH \& Co.
  KG,}}},\ }\href@noop {} {\bibinfo {title} {{{COSMOtherm}}, {{Version
  C3}}.0}},\ \bibinfo {howpublished} {http://www.cosmologic.de}\BibitemShut
  {NoStop}%
\bibitem [{\citenamefont {Pedregosa}\ \emph {et~al.}()\citenamefont
  {Pedregosa}, \citenamefont {Pedregosa}, \citenamefont {Varoquaux},
  \citenamefont {Varoquaux}, \citenamefont {Org}, \citenamefont {Gramfort},
  \citenamefont {Gramfort}, \citenamefont {Michel}, \citenamefont {Michel},
  \citenamefont {Fr}, \citenamefont {Thirion}, \citenamefont {Thirion},
  \citenamefont {Grisel}, \citenamefont {Grisel}, \citenamefont {Blondel},
  \citenamefont {Prettenhofer}, \citenamefont {Prettenhofer}, \citenamefont
  {Weiss}, \citenamefont {Dubourg}, \citenamefont {Dubourg}, \citenamefont
  {Vanderplas}, \citenamefont {Passos}, \citenamefont {Tp},\ and\ \citenamefont
  {Cournapeau}}]{Pedregosa}%
  \BibitemOpen
  \bibfield  {author} {\bibinfo {author} {\bibfnamefont {F.}~\bibnamefont
  {Pedregosa}}, \bibinfo {author} {\bibfnamefont {F.}~\bibnamefont
  {Pedregosa}}, \bibinfo {author} {\bibfnamefont {G.}~\bibnamefont
  {Varoquaux}}, \bibinfo {author} {\bibfnamefont {G.}~\bibnamefont
  {Varoquaux}}, \bibinfo {author} {\bibfnamefont {N.}~\bibnamefont {Org}},
  \bibinfo {author} {\bibfnamefont {A.}~\bibnamefont {Gramfort}}, \bibinfo
  {author} {\bibfnamefont {A.}~\bibnamefont {Gramfort}}, \bibinfo {author}
  {\bibfnamefont {V.}~\bibnamefont {Michel}}, \bibinfo {author} {\bibfnamefont
  {V.}~\bibnamefont {Michel}}, \bibinfo {author} {\bibfnamefont
  {L.}~\bibnamefont {Fr}}, \bibinfo {author} {\bibfnamefont {B.}~\bibnamefont
  {Thirion}}, \bibinfo {author} {\bibfnamefont {B.}~\bibnamefont {Thirion}},
  \bibinfo {author} {\bibfnamefont {O.}~\bibnamefont {Grisel}}, \bibinfo
  {author} {\bibfnamefont {O.}~\bibnamefont {Grisel}}, \bibinfo {author}
  {\bibfnamefont {M.}~\bibnamefont {Blondel}}, \bibinfo {author} {\bibfnamefont
  {P.}~\bibnamefont {Prettenhofer}}, \bibinfo {author} {\bibfnamefont
  {P.}~\bibnamefont {Prettenhofer}}, \bibinfo {author} {\bibfnamefont
  {R.}~\bibnamefont {Weiss}}, \bibinfo {author} {\bibfnamefont
  {V.}~\bibnamefont {Dubourg}}, \bibinfo {author} {\bibfnamefont
  {V.}~\bibnamefont {Dubourg}}, \bibinfo {author} {\bibfnamefont
  {J.}~\bibnamefont {Vanderplas}}, \bibinfo {author} {\bibfnamefont
  {A.}~\bibnamefont {Passos}}, \bibinfo {author} {\bibfnamefont
  {A.}~\bibnamefont {Tp}},\ and\ \bibinfo {author} {\bibfnamefont
  {D.}~\bibnamefont {Cournapeau}},\ }\bibfield  {title} {\bibinfo {title}
  {Scikit-learn: {{Machine Learning}} in {{Python}}},\ }\href@noop {} {\bibinfo
   {journal} {MACHINE LEARNING IN PYTHON}\ }\BibitemShut {NoStop}%
\bibitem [{\citenamefont {Liu}\ \emph {et~al.}(2020)\citenamefont {Liu},
  \citenamefont {Venkatesh}, \citenamefont {McBride}, \citenamefont
  {Reichmanis}, \citenamefont {Meredith},\ and\ \citenamefont
  {Grover}}]{Liu2020}%
  \BibitemOpen
\bibfield  {journal} {  }\bibfield  {author} {\bibinfo {author} {\bibfnamefont
  {A.~L.}\ \bibnamefont {Liu}}, \bibinfo {author} {\bibfnamefont
  {R.}~\bibnamefont {Venkatesh}}, \bibinfo {author} {\bibfnamefont
  {M.}~\bibnamefont {McBride}}, \bibinfo {author} {\bibfnamefont
  {E.}~\bibnamefont {Reichmanis}}, \bibinfo {author} {\bibfnamefont {J.~C.}\
  \bibnamefont {Meredith}},\ and\ \bibinfo {author} {\bibfnamefont {M.~A.}\
  \bibnamefont {Grover}},\ }\bibfield  {title} {\bibinfo {title} {Small {{Data
  Machine Learning}}: {{Classification}} and {{Prediction}} of
  {{Poly}}(ethylene terephthalate) {{Stabilizers Using Molecular
  Descriptors}}},\ }\href {https://doi.org/10.1021/acsapm.0c00921} {\bibfield
  {journal} {\bibinfo  {journal} {ACS Applied Polymer Materials}\ }\textbf
  {\bibinfo {volume} {2}},\ \bibinfo {pages} {5592} (\bibinfo {year}
  {2020})}\BibitemShut {NoStop}%
\bibitem [{\citenamefont {Reshef}\ \emph {et~al.}(2011)\citenamefont {Reshef},
  \citenamefont {Reshef}, \citenamefont {Finucane}, \citenamefont {Grossman},
  \citenamefont {McVean}, \citenamefont {Turnbaugh}, \citenamefont {Lander},
  \citenamefont {Mitzenmacher},\ and\ \citenamefont {Sabeti}}]{Reshef2011}%
  \BibitemOpen
  \bibfield  {author} {\bibinfo {author} {\bibfnamefont {D.~N.}\ \bibnamefont
  {Reshef}}, \bibinfo {author} {\bibfnamefont {Y.~A.}\ \bibnamefont {Reshef}},
  \bibinfo {author} {\bibfnamefont {H.~K.}\ \bibnamefont {Finucane}}, \bibinfo
  {author} {\bibfnamefont {S.~R.}\ \bibnamefont {Grossman}}, \bibinfo {author}
  {\bibfnamefont {G.}~\bibnamefont {McVean}}, \bibinfo {author} {\bibfnamefont
  {P.~J.}\ \bibnamefont {Turnbaugh}}, \bibinfo {author} {\bibfnamefont {E.~S.}\
  \bibnamefont {Lander}}, \bibinfo {author} {\bibfnamefont {M.}~\bibnamefont
  {Mitzenmacher}},\ and\ \bibinfo {author} {\bibfnamefont {P.~C.}\ \bibnamefont
  {Sabeti}},\ }\bibfield  {title} {\bibinfo {title} {Detecting {{Novel
  Associations}} in {{Large Data Sets}}},\ }\href
  {https://doi.org/10.1126/science.1205438} {\bibfield  {journal} {\bibinfo
  {journal} {Science}\ }\textbf {\bibinfo {volume} {334}},\ \bibinfo {pages}
  {1518} (\bibinfo {year} {2011})}\BibitemShut {NoStop}%
\bibitem [{\citenamefont {Luo}\ and\ \citenamefont {Sommer}(2013)}]{Luo2013}%
  \BibitemOpen
  \bibfield  {author} {\bibinfo {author} {\bibfnamefont {C.}~\bibnamefont
  {Luo}}\ and\ \bibinfo {author} {\bibfnamefont {J.-U.}\ \bibnamefont
  {Sommer}},\ }\bibfield  {title} {\bibinfo {title} {Disentanglement of
  {{Linear Polymer Chains Toward Unentangled Crystals}}},\ }\href
  {https://doi.org/10.1021/mz300552x} {\bibfield  {journal} {\bibinfo
  {journal} {ACS Macro Letters}\ }\textbf {\bibinfo {volume} {2}},\ \bibinfo
  {pages} {31} (\bibinfo {year} {2013})}\BibitemShut {NoStop}%
\bibitem [{\citenamefont {Luo}\ \emph {et~al.}(2016)\citenamefont {Luo},
  \citenamefont {Kr{\"o}ger},\ and\ \citenamefont {Sommer}}]{Luo2016}%
  \BibitemOpen
  \bibfield  {author} {\bibinfo {author} {\bibfnamefont {C.}~\bibnamefont
  {Luo}}, \bibinfo {author} {\bibfnamefont {M.}~\bibnamefont {Kr{\"o}ger}},\
  and\ \bibinfo {author} {\bibfnamefont {J.-U.}\ \bibnamefont {Sommer}},\
  }\bibfield  {title} {\bibinfo {title} {Entanglements and {{Crystallization}}
  of {{Concentrated Polymer Solutions}}: {{Molecular Dynamics Simulations}}},\
  }\href {https://doi.org/10.1021/acs.macromol.6b02124} {\bibfield  {journal}
  {\bibinfo  {journal} {Macromolecules}\ }\textbf {\bibinfo {volume} {49}},\
  \bibinfo {pages} {9017} (\bibinfo {year} {2016})}\BibitemShut {NoStop}%
\bibitem [{\citenamefont {Luo}\ and\ \citenamefont {Sommer}(2016)}]{Luo2016a}%
  \BibitemOpen
  \bibfield  {author} {\bibinfo {author} {\bibfnamefont {C.}~\bibnamefont
  {Luo}}\ and\ \bibinfo {author} {\bibfnamefont {J.-U.}\ \bibnamefont
  {Sommer}},\ }\bibfield  {title} {\bibinfo {title} {Role of {{Thermal
  History}} and {{Entanglement Related Thickness Selection}} in {{Polymer
  Crystallization}}},\ }\href {https://doi.org/10.1021/acsmacrolett.5b00668}
  {\bibfield  {journal} {\bibinfo  {journal} {ACS Macro Letters}\ }\textbf
  {\bibinfo {volume} {5}},\ \bibinfo {pages} {30} (\bibinfo {year}
  {2016})}\BibitemShut {NoStop}%
\bibitem [{\citenamefont {Xiao}\ \emph {et~al.}(2017)\citenamefont {Xiao},
  \citenamefont {Luo}, \citenamefont {Yan},\ and\ \citenamefont
  {Sommer}}]{Xiao2017}%
  \BibitemOpen
  \bibfield  {author} {\bibinfo {author} {\bibfnamefont {H.}~\bibnamefont
  {Xiao}}, \bibinfo {author} {\bibfnamefont {C.}~\bibnamefont {Luo}}, \bibinfo
  {author} {\bibfnamefont {D.}~\bibnamefont {Yan}},\ and\ \bibinfo {author}
  {\bibfnamefont {J.-U.}\ \bibnamefont {Sommer}},\ }\bibfield  {title}
  {\bibinfo {title} {Molecular {{Dynamics Simulation}} of {{Crystallization
  Cyclic Polymer Melts As Compared}} to {{Their Linear Counterparts}}},\ }\href
  {https://doi.org/10.1021/acs.macromol.7b01570} {\bibfield  {journal}
  {\bibinfo  {journal} {Macromolecules}\ }\textbf {\bibinfo {volume} {50}},\
  \bibinfo {pages} {9796} (\bibinfo {year} {2017})}\BibitemShut {NoStop}%
\bibitem [{\citenamefont {Yamamoto}(2019)}]{Yamamoto2019}%
  \BibitemOpen
  \bibfield  {author} {\bibinfo {author} {\bibfnamefont {T.}~\bibnamefont
  {Yamamoto}},\ }\bibfield  {title} {\bibinfo {title} {Molecular {{Dynamics
  Simulation}} of {{Stretch-Induced Crystallization}} in {{Polyethylene}}:
  {{Emergence}} of {{Fiber Structure}} and {{Molecular Network}}},\ }\href
  {https://doi.org/10.1021/acs.macromol.8b02569} {\bibfield  {journal}
  {\bibinfo  {journal} {Macromolecules}\ }\textbf {\bibinfo {volume} {52}},\
  \bibinfo {pages} {1695} (\bibinfo {year} {2019})}\BibitemShut {NoStop}%
\bibitem [{\citenamefont {Zhang}\ and\ \citenamefont
  {Larson}(2018)}]{Zhang2018}%
  \BibitemOpen
  \bibfield  {author} {\bibinfo {author} {\bibfnamefont {W.}~\bibnamefont
  {Zhang}}\ and\ \bibinfo {author} {\bibfnamefont {R.~G.}\ \bibnamefont
  {Larson}},\ }\bibfield  {title} {\bibinfo {title} {Direct {{All-Atom
  Molecular Dynamics Simulations}} of the {{Effects}} of {{Short Chain
  Branching}} on {{Polyethylene Oligomer Crystal Nucleation}}},\ }\href
  {https://doi.org/10.1021/acs.macromol.8b00958} {\bibfield  {journal}
  {\bibinfo  {journal} {Macromolecules}\ }\textbf {\bibinfo {volume} {51}},\
  \bibinfo {pages} {4762} (\bibinfo {year} {2018})}\BibitemShut {NoStop}%
\bibitem [{\citenamefont {L{\"u}tzow}\ \emph {et~al.}(1999)\citenamefont
  {L{\"u}tzow}, \citenamefont {Tihminlioglu}, \citenamefont {Danner},
  \citenamefont {Duda}, \citenamefont {De~Haan}, \citenamefont {Warnier},\ and\
  \citenamefont {Zielinski}}]{Lutzow1999}%
  \BibitemOpen
  \bibfield  {author} {\bibinfo {author} {\bibfnamefont {N.}~\bibnamefont
  {L{\"u}tzow}}, \bibinfo {author} {\bibfnamefont {A.}~\bibnamefont
  {Tihminlioglu}}, \bibinfo {author} {\bibfnamefont {R.~P.}\ \bibnamefont
  {Danner}}, \bibinfo {author} {\bibfnamefont {J.~L.}\ \bibnamefont {Duda}},
  \bibinfo {author} {\bibfnamefont {A.}~\bibnamefont {De~Haan}}, \bibinfo
  {author} {\bibfnamefont {G.}~\bibnamefont {Warnier}},\ and\ \bibinfo {author}
  {\bibfnamefont {J.~M.}\ \bibnamefont {Zielinski}},\ }\bibfield  {title}
  {\bibinfo {title} {Diffusion of toluene and n-heptane in polyethylenes of
  different crystallinity},\ }\href
  {https://doi.org/10.1016/S0032-3861(98)00473-X} {\bibfield  {journal}
  {\bibinfo  {journal} {Polymer}\ }\textbf {\bibinfo {volume} {40}},\ \bibinfo
  {pages} {2797} (\bibinfo {year} {1999})}\BibitemShut {NoStop}%
\bibitem [{\citenamefont {Xu}\ \emph {et~al.}(2021)\citenamefont {Xu},
  \citenamefont {Gehui}, \citenamefont {Cao}, \citenamefont {Guo},
  \citenamefont {Serrano}, \citenamefont {Esker},\ and\ \citenamefont
  {Liu}}]{Xu2021}%
  \BibitemOpen
  \bibfield  {author} {\bibinfo {author} {\bibfnamefont {Z.}~\bibnamefont
  {Xu}}, \bibinfo {author} {\bibfnamefont {L.}~\bibnamefont {Gehui}}, \bibinfo
  {author} {\bibfnamefont {K.}~\bibnamefont {Cao}}, \bibinfo {author}
  {\bibfnamefont {D.}~\bibnamefont {Guo}}, \bibinfo {author} {\bibfnamefont
  {J.}~\bibnamefont {Serrano}}, \bibinfo {author} {\bibfnamefont
  {A.}~\bibnamefont {Esker}},\ and\ \bibinfo {author} {\bibfnamefont
  {G.}~\bibnamefont {Liu}},\ }\bibfield  {title} {\bibinfo {title}
  {Solvent-{{Resistant Self-Crosslinked Poly}}(ether imide)},\ }\href
  {https://doi.org/10.1021/acs.macromol.0c02860} {\bibfield  {journal}
  {\bibinfo  {journal} {Macromolecules}\ }\textbf {\bibinfo {volume} {54}},\
  \bibinfo {pages} {3405} (\bibinfo {year} {2021})}\BibitemShut {NoStop}%
\bibitem [{\citenamefont {Rangou}\ \emph {et~al.}(2022)\citenamefont {Rangou},
  \citenamefont {Appold}, \citenamefont {Lademann}, \citenamefont {Buhr},\ and\
  \citenamefont {Filiz}}]{Rangou2022}%
  \BibitemOpen
  \bibfield  {author} {\bibinfo {author} {\bibfnamefont {S.}~\bibnamefont
  {Rangou}}, \bibinfo {author} {\bibfnamefont {M.}~\bibnamefont {Appold}},
  \bibinfo {author} {\bibfnamefont {B.}~\bibnamefont {Lademann}}, \bibinfo
  {author} {\bibfnamefont {K.}~\bibnamefont {Buhr}},\ and\ \bibinfo {author}
  {\bibfnamefont {V.}~\bibnamefont {Filiz}},\ }\bibfield  {title} {\bibinfo
  {title} {Thermally and {{Chemically Stable Isoporous Block Copolymer
  Membranes}}},\ }\href {https://doi.org/10.1021/acsmacrolett.2c00352}
  {\bibfield  {journal} {\bibinfo  {journal} {ACS Macro Letters}\ }\textbf
  {\bibinfo {volume} {11}},\ \bibinfo {pages} {1142} (\bibinfo {year}
  {2022})}\BibitemShut {NoStop}%
\bibitem [{\citenamefont {Ji}\ \emph {et~al.}(2015)\citenamefont {Ji},
  \citenamefont {Wang}, \citenamefont {Yan},\ and\ \citenamefont
  {Wang}}]{Ji2015}%
  \BibitemOpen
  \bibfield  {author} {\bibinfo {author} {\bibfnamefont {X.}~\bibnamefont
  {Ji}}, \bibinfo {author} {\bibfnamefont {Z.}~\bibnamefont {Wang}}, \bibinfo
  {author} {\bibfnamefont {J.}~\bibnamefont {Yan}},\ and\ \bibinfo {author}
  {\bibfnamefont {Z.}~\bibnamefont {Wang}},\ }\bibfield  {title} {\bibinfo
  {title} {Partially bio-based polyimides from isohexide-derived diamines},\
  }\href {https://doi.org/10.1016/j.polymer.2015.07.051} {\bibfield  {journal}
  {\bibinfo  {journal} {Polymer}\ }\textbf {\bibinfo {volume} {74}},\ \bibinfo
  {pages} {38} (\bibinfo {year} {2015})}\BibitemShut {NoStop}%
\bibitem [{\citenamefont {Du}\ \emph {et~al.}(2013)\citenamefont {Du},
  \citenamefont {Zhang}, \citenamefont {Bai},\ and\ \citenamefont
  {Li}}]{Du2013}%
  \BibitemOpen
  \bibfield  {author} {\bibinfo {author} {\bibfnamefont {C.}~\bibnamefont
  {Du}}, \bibinfo {author} {\bibfnamefont {A.}~\bibnamefont {Zhang}}, \bibinfo
  {author} {\bibfnamefont {H.}~\bibnamefont {Bai}},\ and\ \bibinfo {author}
  {\bibfnamefont {L.}~\bibnamefont {Li}},\ }\bibfield  {title} {\bibinfo
  {title} {Robust {{Microsieves}} with {{Excellent Solvent Resistance}}:
  {{Cross-Linkage}} of {{Perforated Polymer Films}} with {{Honeycomb
  Structure}}},\ }\href {https://doi.org/10.1021/mz300616z} {\bibfield
  {journal} {\bibinfo  {journal} {ACS Macro Letters}\ }\textbf {\bibinfo
  {volume} {2}},\ \bibinfo {pages} {27} (\bibinfo {year} {2013})}\BibitemShut
  {NoStop}%
\bibitem [{\citenamefont {Dugas}\ \emph {et~al.}(2023)\citenamefont {Dugas},
  \citenamefont {Zhong}, \citenamefont {Park}, \citenamefont {Jiang},
  \citenamefont {Ouimet}, \citenamefont {Xu}, \citenamefont {Schaefer},\ and\
  \citenamefont {Phillip}}]{Dugas2023}%
  \BibitemOpen
  \bibfield  {author} {\bibinfo {author} {\bibfnamefont {M.~P.}\ \bibnamefont
  {Dugas}}, \bibinfo {author} {\bibfnamefont {S.}~\bibnamefont {Zhong}},
  \bibinfo {author} {\bibfnamefont {B.}~\bibnamefont {Park}}, \bibinfo {author}
  {\bibfnamefont {J.}~\bibnamefont {Jiang}}, \bibinfo {author} {\bibfnamefont
  {J.~A.}\ \bibnamefont {Ouimet}}, \bibinfo {author} {\bibfnamefont
  {J.}~\bibnamefont {Xu}}, \bibinfo {author} {\bibfnamefont {J.~L.}\
  \bibnamefont {Schaefer}},\ and\ \bibinfo {author} {\bibfnamefont {W.~A.}\
  \bibnamefont {Phillip}},\ }\bibfield  {title} {\bibinfo {title} {Influence of
  {{Solvent Affinity}} on {{Transport}} through {{Cross-Linked Copolymer
  Membranes}} for {{Organic Solvent Nanofiltration}}},\ }\href
  {https://doi.org/10.1021/acsapm.3c00460} {\bibfield  {journal} {\bibinfo
  {journal} {ACS Applied Polymer Materials}\ }\textbf {\bibinfo {volume} {5}},\
  \bibinfo {pages} {6781} (\bibinfo {year} {2023})}\BibitemShut {NoStop}%
\bibitem [{\citenamefont {Toolan}\ \emph {et~al.}(2016)\citenamefont {Toolan},
  \citenamefont {Isakova}, \citenamefont {Hodgkinson}, \citenamefont
  {{Reeves-McLaren}}, \citenamefont {Hammond}, \citenamefont {Edler},
  \citenamefont {Briscoe}, \citenamefont {Arnold}, \citenamefont {Gough},
  \citenamefont {Topham},\ and\ \citenamefont {Howse}}]{Toolan2016}%
  \BibitemOpen
  \bibfield  {author} {\bibinfo {author} {\bibfnamefont {D.~T.~W.}\
  \bibnamefont {Toolan}}, \bibinfo {author} {\bibfnamefont {A.}~\bibnamefont
  {Isakova}}, \bibinfo {author} {\bibfnamefont {R.}~\bibnamefont {Hodgkinson}},
  \bibinfo {author} {\bibfnamefont {N.}~\bibnamefont {{Reeves-McLaren}}},
  \bibinfo {author} {\bibfnamefont {O.~S.}\ \bibnamefont {Hammond}}, \bibinfo
  {author} {\bibfnamefont {K.~J.}\ \bibnamefont {Edler}}, \bibinfo {author}
  {\bibfnamefont {W.~H.}\ \bibnamefont {Briscoe}}, \bibinfo {author}
  {\bibfnamefont {T.}~\bibnamefont {Arnold}}, \bibinfo {author} {\bibfnamefont
  {T.}~\bibnamefont {Gough}}, \bibinfo {author} {\bibfnamefont {P.~D.}\
  \bibnamefont {Topham}},\ and\ \bibinfo {author} {\bibfnamefont {J.~R.}\
  \bibnamefont {Howse}},\ }\bibfield  {title} {\bibinfo {title} {Insights into
  the {{Influence}} of {{Solvent Polarity}} on the {{Crystallization}} of
  {{Poly}}(ethylene oxide) {{Spin-Coated Thin Films}} via in {{Situ Grazing
  Incidence Wide-Angle X-ray Scattering}}},\ }\href
  {https://doi.org/10.1021/acs.macromol.6b00312} {\bibfield  {journal}
  {\bibinfo  {journal} {Macromolecules}\ }\textbf {\bibinfo {volume} {49}},\
  \bibinfo {pages} {4579} (\bibinfo {year} {2016})}\BibitemShut {NoStop}%
\bibitem [{\citenamefont {Lang}\ \emph {et~al.}(2022)\citenamefont {Lang},
  \citenamefont {Scholz}, \citenamefont {L{\"o}ser}, \citenamefont {Bunk},
  \citenamefont {Fribiczer}, \citenamefont {Seiffert}, \citenamefont
  {B{\"o}hme},\ and\ \citenamefont {Saalw{\"a}chter}}]{Lang2022}%
  \BibitemOpen
  \bibfield  {author} {\bibinfo {author} {\bibfnamefont {M.}~\bibnamefont
  {Lang}}, \bibinfo {author} {\bibfnamefont {R.}~\bibnamefont {Scholz}},
  \bibinfo {author} {\bibfnamefont {L.}~\bibnamefont {L{\"o}ser}}, \bibinfo
  {author} {\bibfnamefont {C.}~\bibnamefont {Bunk}}, \bibinfo {author}
  {\bibfnamefont {N.}~\bibnamefont {Fribiczer}}, \bibinfo {author}
  {\bibfnamefont {S.}~\bibnamefont {Seiffert}}, \bibinfo {author}
  {\bibfnamefont {F.}~\bibnamefont {B{\"o}hme}},\ and\ \bibinfo {author}
  {\bibfnamefont {K.}~\bibnamefont {Saalw{\"a}chter}},\ }\bibfield  {title}
  {\bibinfo {title} {Swelling and {{Residual Bond Orientations}} of {{Polymer
  Model Gels}}: {{The Entanglement-Free Limit}}},\ }\href
  {https://doi.org/10.1021/acs.macromol.2c00589} {\bibfield  {journal}
  {\bibinfo  {journal} {Macromolecules}\ }\textbf {\bibinfo {volume} {55}},\
  \bibinfo {pages} {5997} (\bibinfo {year} {2022})}\BibitemShut {NoStop}%
\bibitem [{\citenamefont {Navarro}\ \emph {et~al.}(2022)\citenamefont
  {Navarro}, \citenamefont {Th{\"u}nemann},\ and\ \citenamefont
  {Klinger}}]{Navarro2022}%
  \BibitemOpen
  \bibfield  {author} {\bibinfo {author} {\bibfnamefont {L.}~\bibnamefont
  {Navarro}}, \bibinfo {author} {\bibfnamefont {A.~F.}\ \bibnamefont
  {Th{\"u}nemann}},\ and\ \bibinfo {author} {\bibfnamefont {D.}~\bibnamefont
  {Klinger}},\ }\bibfield  {title} {\bibinfo {title} {Solvent {{Annealing}} of
  {{Striped Ellipsoidal Block Copolymer Particles}}: {{Reversible Control}}
  over {{Lamellae Asymmetry}}, {{Aspect Ratio}}, and {{Particle Surface}}},\
  }\href {https://doi.org/10.1021/acsmacrolett.1c00665} {\bibfield  {journal}
  {\bibinfo  {journal} {ACS Macro Letters}\ }\textbf {\bibinfo {volume} {11}},\
  \bibinfo {pages} {329} (\bibinfo {year} {2022})}\BibitemShut {NoStop}%
\bibitem [{\citenamefont {Nezili}\ \emph {et~al.}(2023)\citenamefont {Nezili},
  \citenamefont {Mdarhri}, \citenamefont {El~Aboudi}, \citenamefont {Brosseau},
  \citenamefont {Zaghrioui}, \citenamefont {Ghorbal}, \citenamefont {He},\ and\
  \citenamefont {Bai}}]{Nezili2023}%
  \BibitemOpen
  \bibfield  {author} {\bibinfo {author} {\bibfnamefont {Y.}~\bibnamefont
  {Nezili}}, \bibinfo {author} {\bibfnamefont {A.}~\bibnamefont {Mdarhri}},
  \bibinfo {author} {\bibfnamefont {I.}~\bibnamefont {El~Aboudi}}, \bibinfo
  {author} {\bibfnamefont {C.}~\bibnamefont {Brosseau}}, \bibinfo {author}
  {\bibfnamefont {M.}~\bibnamefont {Zaghrioui}}, \bibinfo {author}
  {\bibfnamefont {A.}~\bibnamefont {Ghorbal}}, \bibinfo {author} {\bibfnamefont
  {D.}~\bibnamefont {He}},\ and\ \bibinfo {author} {\bibfnamefont
  {J.}~\bibnamefont {Bai}},\ }\bibfield  {title} {\bibinfo {title} {Solvent
  polarity impacts the sorption kinetics and tensile properties of carbon black
  filled elastomers},\ }\href {https://doi.org/10.1016/j.polymer.2022.125563}
  {\bibfield  {journal} {\bibinfo  {journal} {Polymer}\ }\textbf {\bibinfo
  {volume} {264}},\ \bibinfo {pages} {125563} (\bibinfo {year}
  {2023})}\BibitemShut {NoStop}%
\bibitem [{\citenamefont {Jorgensen}\ \emph {et~al.}(1996)\citenamefont
  {Jorgensen}, \citenamefont {Maxwell},\ and\ \citenamefont
  {{Tirado-Rives}}}]{Jorgensen1996}%
  \BibitemOpen
  \bibfield  {author} {\bibinfo {author} {\bibfnamefont {W.~L.}\ \bibnamefont
  {Jorgensen}}, \bibinfo {author} {\bibfnamefont {D.~S.}\ \bibnamefont
  {Maxwell}},\ and\ \bibinfo {author} {\bibfnamefont {J.}~\bibnamefont
  {{Tirado-Rives}}},\ }\bibfield  {title} {\bibinfo {title} {Development and
  {{Testing}} of the {{OPLS All-Atom Force Field}} on {{Conformational
  Energetics}} and {{Properties}} of {{Organic Liquids}}},\ }\href
  {https://doi.org/10.1021/ja9621760} {\bibfield  {journal} {\bibinfo
  {journal} {Journal of the American Chemical Society}\ }\textbf {\bibinfo
  {volume} {118}},\ \bibinfo {pages} {11225} (\bibinfo {year}
  {1996})}\BibitemShut {NoStop}%
\bibitem [{\citenamefont {Jorge}\ \emph {et~al.}(2021)\citenamefont {Jorge},
  \citenamefont {Milne}, \citenamefont {Barrera},\ and\ \citenamefont
  {Gomes}}]{Jorge2021}%
  \BibitemOpen
  \bibfield  {author} {\bibinfo {author} {\bibfnamefont {M.}~\bibnamefont
  {Jorge}}, \bibinfo {author} {\bibfnamefont {A.~W.}\ \bibnamefont {Milne}},
  \bibinfo {author} {\bibfnamefont {M.~C.}\ \bibnamefont {Barrera}},\ and\
  \bibinfo {author} {\bibfnamefont {J.~R.~B.}\ \bibnamefont {Gomes}},\
  }\bibfield  {title} {\bibinfo {title} {New {{Force-Field}} for
  {{Organosilicon Molecules}} in the {{Liquid Phase}}},\ }\href
  {https://doi.org/10.1021/acsphyschemau.1c00014} {\bibfield  {journal}
  {\bibinfo  {journal} {ACS Physical Chemistry Au}\ }\textbf {\bibinfo {volume}
  {1}},\ \bibinfo {pages} {54} (\bibinfo {year} {2021})}\BibitemShut {NoStop}%
\bibitem [{\citenamefont {Venezia}\ \emph {et~al.}(2025)\citenamefont
  {Venezia}, \citenamefont {Correa}, \citenamefont {Esposito}, \citenamefont
  {Muna{\`o}}, \citenamefont {Nicola},\ and\ \citenamefont
  {Milano}}]{Venezia2025}%
  \BibitemOpen
  \bibfield  {author} {\bibinfo {author} {\bibfnamefont {E.}~\bibnamefont
  {Venezia}}, \bibinfo {author} {\bibfnamefont {A.}~\bibnamefont {Correa}},
  \bibinfo {author} {\bibfnamefont {R.}~\bibnamefont {Esposito}}, \bibinfo
  {author} {\bibfnamefont {G.}~\bibnamefont {Muna{\`o}}}, \bibinfo {author}
  {\bibfnamefont {A.~D.}\ \bibnamefont {Nicola}},\ and\ \bibinfo {author}
  {\bibfnamefont {G.}~\bibnamefont {Milano}},\ }\bibfield  {title} {\bibinfo
  {title} {Molecular {{Models}} of {{Nanoplastics}} from {{Semi-Crystalline
  Polyethylene}}},\ }\href {https://doi.org/10.1021/acs.macromol.4c03240}
  {\bibfield  {journal} {\bibinfo  {journal} {Macromolecules}\ }\textbf
  {\bibinfo {volume} {58}},\ \bibinfo {pages} {3119} (\bibinfo {year}
  {2025})}\BibitemShut {NoStop}%
\bibitem [{\citenamefont {Pawlak}\ and\ \citenamefont
  {Galeski}(2008)}]{Pawlak2008}%
  \BibitemOpen
  \bibfield  {author} {\bibinfo {author} {\bibfnamefont {A.}~\bibnamefont
  {Pawlak}}\ and\ \bibinfo {author} {\bibfnamefont {A.}~\bibnamefont
  {Galeski}},\ }\bibfield  {title} {\bibinfo {title} {Cavitation during
  {{Tensile Deformation}} of {{Polypropylene}}},\ }\href
  {https://doi.org/10.1021/ma0715122} {\bibfield  {journal} {\bibinfo
  {journal} {Macromolecules}\ }\textbf {\bibinfo {volume} {41}},\ \bibinfo
  {pages} {2839} (\bibinfo {year} {2008})}\BibitemShut {NoStop}%
\bibitem [{\citenamefont {Puleo}\ \emph {et~al.}(1989)\citenamefont {Puleo},
  \citenamefont {Paul},\ and\ \citenamefont {Wong}}]{Puleo1989}%
  \BibitemOpen
  \bibfield  {author} {\bibinfo {author} {\bibfnamefont {A.~C.}\ \bibnamefont
  {Puleo}}, \bibinfo {author} {\bibfnamefont {D.~R.}\ \bibnamefont {Paul}},\
  and\ \bibinfo {author} {\bibfnamefont {P.~K.}\ \bibnamefont {Wong}},\
  }\bibfield  {title} {\bibinfo {title} {Gas sorption and transport in
  semicrystalline poly(4-methyl-1-pentene)},\ }\href
  {https://doi.org/10.1016/0032-3861(89)90060-8} {\bibfield  {journal}
  {\bibinfo  {journal} {Polymer}\ }\textbf {\bibinfo {volume} {30}},\ \bibinfo
  {pages} {1357} (\bibinfo {year} {1989})}\BibitemShut {NoStop}%
\bibitem [{\citenamefont {Thomas}\ and\ \citenamefont
  {Windle}(1982)}]{Thomas1982}%
  \BibitemOpen
  \bibfield  {author} {\bibinfo {author} {\bibfnamefont {N.~L.}\ \bibnamefont
  {Thomas}}\ and\ \bibinfo {author} {\bibfnamefont {A.~H.}\ \bibnamefont
  {Windle}},\ }\bibfield  {title} {\bibinfo {title} {A theory of case {{II}}
  diffusion},\ }\href {https://doi.org/10.1016/0032-3861(82)90093-3} {\bibfield
   {journal} {\bibinfo  {journal} {Polymer}\ }\textbf {\bibinfo {volume}
  {23}},\ \bibinfo {pages} {529} (\bibinfo {year} {1982})}\BibitemShut
  {NoStop}%
\bibitem [{\citenamefont {Peppas}\ \emph {et~al.}(1994)\citenamefont {Peppas},
  \citenamefont {Wu},\ and\ \citenamefont {{von Meerwall}}}]{Peppas1994}%
  \BibitemOpen
  \bibfield  {author} {\bibinfo {author} {\bibfnamefont {N.~A.}\ \bibnamefont
  {Peppas}}, \bibinfo {author} {\bibfnamefont {J.~C.}\ \bibnamefont {Wu}},\
  and\ \bibinfo {author} {\bibfnamefont {E.~D.}\ \bibnamefont {{von
  Meerwall}}},\ }\bibfield  {title} {\bibinfo {title} {Mathematical
  {{Modeling}} and {{Experimental Characterization}} of {{Polymer
  Dissolution}}},\ }\href {https://doi.org/10.1021/ma00098a017} {\bibfield
  {journal} {\bibinfo  {journal} {Macromolecules}\ }\textbf {\bibinfo {volume}
  {27}},\ \bibinfo {pages} {5626} (\bibinfo {year} {1994})}\BibitemShut
  {NoStop}%
\bibitem [{\citenamefont {Devotta}\ \emph {et~al.}(1994)\citenamefont
  {Devotta}, \citenamefont {Premnath}, \citenamefont {Badiger}, \citenamefont
  {Rajamohanan}, \citenamefont {Ganapathy},\ and\ \citenamefont
  {Mashelkar}}]{Devotta1994}%
  \BibitemOpen
  \bibfield  {author} {\bibinfo {author} {\bibfnamefont {I.}~\bibnamefont
  {Devotta}}, \bibinfo {author} {\bibfnamefont {V.}~\bibnamefont {Premnath}},
  \bibinfo {author} {\bibfnamefont {M.~V.}\ \bibnamefont {Badiger}}, \bibinfo
  {author} {\bibfnamefont {P.~R.}\ \bibnamefont {Rajamohanan}}, \bibinfo
  {author} {\bibfnamefont {S.}~\bibnamefont {Ganapathy}},\ and\ \bibinfo
  {author} {\bibfnamefont {R.~A.}\ \bibnamefont {Mashelkar}},\ }\bibfield
  {title} {\bibinfo {title} {On the dynamics of mobilization in
  swelling-dissolving polymeric systems},\ }\href
  {https://doi.org/10.1021/ma00080a030} {\bibfield  {journal} {\bibinfo
  {journal} {Macromolecules}\ }\textbf {\bibinfo {volume} {27}},\ \bibinfo
  {pages} {532} (\bibinfo {year} {1994})}\BibitemShut {NoStop}%
\bibitem [{\citenamefont {Gardeniers}\ \emph {et~al.}(2022)\citenamefont
  {Gardeniers}, \citenamefont {Mani}, \citenamefont {{de Boer}}, \citenamefont
  {{Hermida-Merino}}, \citenamefont {Graf}, \citenamefont {Rastogi},\ and\
  \citenamefont {Harings}}]{Gardeniers2022}%
  \BibitemOpen
  \bibfield  {author} {\bibinfo {author} {\bibfnamefont {M.}~\bibnamefont
  {Gardeniers}}, \bibinfo {author} {\bibfnamefont {M.}~\bibnamefont {Mani}},
  \bibinfo {author} {\bibfnamefont {E.}~\bibnamefont {{de Boer}}}, \bibinfo
  {author} {\bibfnamefont {D.}~\bibnamefont {{Hermida-Merino}}}, \bibinfo
  {author} {\bibfnamefont {R.}~\bibnamefont {Graf}}, \bibinfo {author}
  {\bibfnamefont {S.}~\bibnamefont {Rastogi}},\ and\ \bibinfo {author}
  {\bibfnamefont {J.~A.~W.}\ \bibnamefont {Harings}},\ }\bibfield  {title}
  {\bibinfo {title} {Hydration, {{Refinement}}, and {{Dissolution}} of the
  {{Crystalline Phase}} in {{Polyamide}} 6 {{Polymorphs}} for {{Ultimate
  Thermomechanical Properties}}},\ }\href
  {https://doi.org/10.1021/acs.macromol.2c00211} {\bibfield  {journal}
  {\bibinfo  {journal} {Macromolecules}\ }\textbf {\bibinfo {volume} {55}},\
  \bibinfo {pages} {5080} (\bibinfo {year} {2022})}\BibitemShut {NoStop}%
\bibitem [{\citenamefont {Ribar}\ \emph {et~al.}(2000)\citenamefont {Ribar},
  \citenamefont {Bhargava},\ and\ \citenamefont {Koenig}}]{Ribar2000}%
  \BibitemOpen
  \bibfield  {author} {\bibinfo {author} {\bibfnamefont {T.}~\bibnamefont
  {Ribar}}, \bibinfo {author} {\bibfnamefont {R.}~\bibnamefont {Bhargava}},\
  and\ \bibinfo {author} {\bibfnamefont {J.~L.}\ \bibnamefont {Koenig}},\
  }\bibfield  {title} {\bibinfo {title} {{{FT-IR Imaging}} of {{Polymer
  Dissolution}} by {{Solvent Mixtures}}. 1. {{Solvents}}},\ }\href
  {https://doi.org/10.1021/ma000851r} {\bibfield  {journal} {\bibinfo
  {journal} {Macromolecules}\ }\textbf {\bibinfo {volume} {33}},\ \bibinfo
  {pages} {8842} (\bibinfo {year} {2000})}\BibitemShut {NoStop}%
\bibitem [{\citenamefont {Keller}(1957)}]{Keller1957}%
  \BibitemOpen
  \bibfield  {author} {\bibinfo {author} {\bibfnamefont {A.}~\bibnamefont
  {Keller}},\ }\bibfield  {title} {\bibinfo {title} {A note on single crystals
  in polymers: {{Evidence}} for a folded chain configuration},\ }\href
  {https://doi.org/10.1080/14786435708242746} {\bibfield  {journal} {\bibinfo
  {journal} {The Philosophical Magazine: A Journal of Theoretical Experimental
  and Applied Physics}\ }\textbf {\bibinfo {volume} {2}},\ \bibinfo {pages}
  {1171} (\bibinfo {year} {1957})}\BibitemShut {NoStop}%
\bibitem [{\citenamefont {Kanomi}\ \emph {et~al.}(2023)\citenamefont {Kanomi},
  \citenamefont {Marubayashi}, \citenamefont {Miyata},\ and\ \citenamefont
  {Jinnai}}]{Kanomi2023}%
  \BibitemOpen
  \bibfield  {author} {\bibinfo {author} {\bibfnamefont {S.}~\bibnamefont
  {Kanomi}}, \bibinfo {author} {\bibfnamefont {H.}~\bibnamefont {Marubayashi}},
  \bibinfo {author} {\bibfnamefont {T.}~\bibnamefont {Miyata}},\ and\ \bibinfo
  {author} {\bibfnamefont {H.}~\bibnamefont {Jinnai}},\ }\bibfield  {title}
  {\bibinfo {title} {Reassessing chain tilt in the lamellar crystals of
  polyethylene},\ }\href {https://doi.org/10.1038/s41467-023-41138-4}
  {\bibfield  {journal} {\bibinfo  {journal} {Nature Communications}\ }\textbf
  {\bibinfo {volume} {14}},\ \bibinfo {pages} {5531} (\bibinfo {year}
  {2023})}\BibitemShut {NoStop}%
\bibitem [{\citenamefont {Ercken}\ \emph {et~al.}(1996)\citenamefont {Ercken},
  \citenamefont {Adriaensens}, \citenamefont {Reggers}, \citenamefont
  {Carleer}, \citenamefont {Vanderzande},\ and\ \citenamefont
  {Gelan}}]{Ercken1996}%
  \BibitemOpen
  \bibfield  {author} {\bibinfo {author} {\bibfnamefont {M.}~\bibnamefont
  {Ercken}}, \bibinfo {author} {\bibfnamefont {P.}~\bibnamefont {Adriaensens}},
  \bibinfo {author} {\bibfnamefont {G.}~\bibnamefont {Reggers}}, \bibinfo
  {author} {\bibfnamefont {R.}~\bibnamefont {Carleer}}, \bibinfo {author}
  {\bibfnamefont {D.}~\bibnamefont {Vanderzande}},\ and\ \bibinfo {author}
  {\bibfnamefont {J.}~\bibnamefont {Gelan}},\ }\bibfield  {title} {\bibinfo
  {title} {Use of {{Magnetic Resonance Imaging To Study Transport}} of
  {{Methanol}} in {{Poly}}(methyl methacrylate) at {{Variable Temperature}}},\
  }\href {https://doi.org/10.1021/ma960155k} {\bibfield  {journal} {\bibinfo
  {journal} {Macromolecules}\ }\textbf {\bibinfo {volume} {29}},\ \bibinfo
  {pages} {5671} (\bibinfo {year} {1996})}\BibitemShut {NoStop}%
\bibitem [{\citenamefont {Narasimhan}\ and\ \citenamefont
  {Peppas}(1996)}]{Narasimhan1996}%
  \BibitemOpen
  \bibfield  {author} {\bibinfo {author} {\bibfnamefont {B.}~\bibnamefont
  {Narasimhan}}\ and\ \bibinfo {author} {\bibfnamefont {N.~A.}\ \bibnamefont
  {Peppas}},\ }\bibfield  {title} {\bibinfo {title} {On the {{Importance}} of
  {{Chain Reptation}} in {{Models}} of {{Dissolution}} of {{Glassy
  Polymers}}},\ }\href {https://doi.org/10.1021/ma951450s} {\bibfield
  {journal} {\bibinfo  {journal} {Macromolecules}\ }\textbf {\bibinfo {volume}
  {29}},\ \bibinfo {pages} {3283} (\bibinfo {year} {1996})}\BibitemShut
  {NoStop}%
\bibitem [{\citenamefont {Lyu}\ \emph {et~al.}(2024)\citenamefont {Lyu},
  \citenamefont {Ding}, \citenamefont {Doi},\ and\ \citenamefont
  {Man}}]{Lyu2024}%
  \BibitemOpen
  \bibfield  {author} {\bibinfo {author} {\bibfnamefont {P.}~\bibnamefont
  {Lyu}}, \bibinfo {author} {\bibfnamefont {Z.}~\bibnamefont {Ding}}, \bibinfo
  {author} {\bibfnamefont {M.}~\bibnamefont {Doi}},\ and\ \bibinfo {author}
  {\bibfnamefont {X.}~\bibnamefont {Man}},\ }\bibfield  {title} {\bibinfo
  {title} {A {{Unified Model}} for {{Non-Fickian Diffusion}} and {{Anomalous
  Swelling}} of {{Glassy Polymer Gels}}},\ }\href
  {https://doi.org/10.1021/acsmacrolett.4c00041} {\bibfield  {journal}
  {\bibinfo  {journal} {ACS Macro Letters}\ }\textbf {\bibinfo {volume} {13}},\
  \bibinfo {pages} {483} (\bibinfo {year} {2024})}\BibitemShut {NoStop}%
\bibitem [{\citenamefont {Brown}\ and\ \citenamefont
  {Dattelbaum}(2005)}]{Brown2005}%
  \BibitemOpen
  \bibfield  {author} {\bibinfo {author} {\bibfnamefont {E.~N.}\ \bibnamefont
  {Brown}}\ and\ \bibinfo {author} {\bibfnamefont {D.~M.}\ \bibnamefont
  {Dattelbaum}},\ }\bibfield  {title} {\bibinfo {title} {The role of
  crystalline phase on fracture and microstructure evolution of
  polytetrafluoroethylene ({{PTFE}})},\ }\href
  {https://doi.org/10.1016/j.polymer.2005.01.061} {\bibfield  {journal}
  {\bibinfo  {journal} {Polymer}\ }\textbf {\bibinfo {volume} {46}},\ \bibinfo
  {pages} {3056} (\bibinfo {year} {2005})}\BibitemShut {NoStop}%
\end{thebibliography}%

\end{document}